\documentclass[aps,twocolumn,a4paper,showkeys,nofootinbib,longbibliography]{revtex4-1}
\usepackage[utf8]{inputenc}
\usepackage{filecontents}
\usepackage{natbib}
\usepackage{amsmath,amssymb,bm,amsthm}
\usepackage{subfigure}
\usepackage{graphicx}
\usepackage{dcolumn}
\usepackage{bm}
\usepackage[mathlines]{lineno}
\usepackage{booktabs}
\usepackage{color}
\newcounter{one}
\usepackage{url}

\usepackage{textcase}

\usepackage{colortbl}
\usepackage{tabularx}
\usepackage{verbatim}
\usepackage{multirow}

\usepackage[T1]{fontenc} 
\usepackage{lmodern}
\usepackage{bbm}
\usepackage[utf8]{inputenc}
\usepackage{amsfonts}
\usepackage{array}

\usepackage{bbm}
\usepackage{enumitem}
\usepackage{umoline}
\usepackage[usenames,svgnames]{xcolor}
\usepackage{natbib}
\usepackage[hyperindex,breaklinks]{hyperref}
\hypersetup{
     colorlinks=true,       		
     linkcolor=Navy,          	
     citecolor=Navy,            
     filecolor=Navy,      		
     urlcolor=Navy,           	
    runcolor=cyan,
 }
\usepackage{graphicx}
\usepackage{amsfonts}
\usepackage{amssymb}
\usepackage{amsmath}
\usepackage{bbm}
\usepackage{enumitem}

\newcommand{\tr}[0]{ {\rm tr}}

\newcommand{\half}[1]{{ \rm h}}
\newcommand{\Oorderof}{\mathcal{O}}
\newcommand{\orderof}[1]{\Oorderof(#1)} 

\newcommand{\for}[0]{\quad \textrm{for} \quad}

\newcommand{\dist}{d}

\newcommand{\co}{{\rm c}}

\newcommand{\diam}{{\rm diam}}

\newcommand{\Set}{\mathcal{S}}

\newcommand{\ad}{{\rm ad}}

\def\beq{\begin{equation}}
\def\eeq{\end{equation}}
\def\nbeq{\begin{equation*}}
\def\neeq{\end{equation*}}
\def\<{\langle}
\def\>{\rangle}

\def\tr{{\rm tr}}

\newcommand{\ed}{X}

\newcommand{\Pau}{{\rm P}}

\newtheorem{theorem}{Theorem}

\newtheorem{lemma}[theorem]{Lemma}
\newtheorem{corol}[theorem]{Corollary}
\newtheorem{assump}{Assumption} 

\newtheorem{prop}[theorem]{Proposition} 

\newcommand{\bal}[2]{#1[#2]}

\newcommand{\sectionprl}[1]{{\par\it #1.---}}

\newcommand{\br}[1]{\left( #1 \right)}
 \newcommand{\norm}[1]{\left \|  #1 \right \|}
  \newcommand{\normb}[1]{\bigl \|  #1 \bigr \|}
\newcommand{\ftr}[0]{\tilde{{\rm tr} }}

\usepackage{mathtools}
\def\multiset#1#2{\ensuremath{\left(\kern-.3em\left(\genfrac{}{}{0pt}{}{#1}{#2}\right)\kern-.3em\right)}}

\usepackage{minitoc}
\usepackage[toc,page,titletoc]{appendix}
\usepackage[capitalise,nameinlink]{cleveref}

\crefname{supp}{Supplement}{Supplements}

\begin{document}
\title{Absence of Fast Scrambling in Thermodynamically Stable Long-Range \\
Interacting Systems} 

\author{Tomotaka Kuwahara$^{1,2}$ and Keiji Saito$^{3}$}
\affiliation{$^{1}$
Mathematical Science Team, RIKEN Center for Advanced Intelligence Project (AIP),1-4-1 Nihonbashi, Chuo-ku, Tokyo 103-0027, Japan
}
\affiliation{$^{2}$Interdisciplinary Theoretical \& Mathematical Sciences Program (iTHEMS) RIKEN 2-1, Hirosawa, Wako, Saitama 351-0198, Japan}

\affiliation{$^{3}$Department of Physics, Keio University, Yokohama 223-8522, Japan}

\begin{abstract}

In this study, we investigate out-of-time-order correlators (OTOCs) in systems with power-law decaying interactions such as $R^{-\alpha}$, where $R$ is the distance. In such systems, the fast scrambling of quantum information or the exponential growth of information propagation can potentially occur according to the decay rate $\alpha$. In this regard,  a crucial open challenge is to identify the optimal condition for $\alpha$ such that fast scrambling cannot occur. In this study, we disprove fast scrambling in generic long-range interacting systems with $\alpha>D$ ($D$: spatial dimension), where the total energy is extensive in terms of system size and the thermodynamic limit is well-defined. We rigorously demonstrate that the OTOC shows a polynomial growth over time as long as $\alpha>D$ and the necessary scrambling time over a distance $R$ is larger than $t\gtrsim R^{\frac{2\alpha-2D}{2\alpha-D+1}}$.

\end{abstract}

\maketitle

\sectionprl{Introduction}
Information scrambling, which characterizes the inaccessibility of local information after time evolution, is a central research topic 
in interdisciplinary problems ranging from thermalization in quantum many-body systems~\cite{PhysRevA.43.2046,PhysRevE.50.888,PhysRevLett.80.1373,Popescu2006} to the black hole information problem~\cite{Hayden_2007,Harrow2009,Lashkari2013}. 
In the recent developments on the connection between quantum chaos and information theory, out-of-time-order correlators (OTOCs) were found to be a useful quantitative tool for characterizing information scrambling~\cite{larkin1969quasiclassical,kitaev2014hidden,Maldacena2016,Swingle2018}.

For quantum lattice models, the OTOC has the form ~\cite{Swingle2018}
\begin{align}
\label{OTOC_def}
C(R,t) :=\frac{1}{\tr (\hat{1})} \tr ([W_i(t),V_{i'}]^\dagger [W_i(t),V_{i'}] ) ,
\end{align}
where $W_i(t)=e^{iHt} W_i e^{-iHt}$, $H$ denotes the system Hamiltonian, and the operators $W_i$ and $V_{i'}$ are defined on the sites $i$ and $i'$, respectively; they are separated from each other by a distance $R$. When the Hamiltonian $H$ includes only short-range interactions, the OTOC grows as 
$C(R,t) \propto e^{\lambda_L (t - R/v_B)}$, where $\lambda_L$ and $v_B$ are referred to as quantum analogs of the Lyapunov exponent~\cite{Maldacena2016} and the butterfly speed~\cite{PhysRevLett.117.091602}, respectively. On the butterfly speed $v_B$, the Lieb--Robinson bound~\cite{ref:LR-bound72,PhysRevLett.97.050401,ref:Nachtergaele2006-LR} yields the simplest upper bound for generic quantum many-body systems. The exploration of the universal behaviors of the OTOC has been one of the most fascinating and essential topics in modern physics~\cite{PhysRevX.7.031016,PhysRevA.95.012120,PhysRevB.98.144304,PhysRevX.8.021014,PhysRevX.8.021013,PhysRevX.9.041017,PhysRevX.9.031048,PhysRevLett.123.010601,PhysRevB.100.195107,PhysRevB.99.224305,Xu2020}. Moreover, along with theoretical developments, the experimental observations of the OTOC have been proposed and realized in various setups~\cite{PhysRevA.94.040302,Garttner2017,PhysRevX.7.031011,PhysRevX.9.021061,Landsman2019,PhysRevLett.124.240505}.

When the Hamiltonian consists of only short-range interactions, the OTOC exhibits a ballistic spreading of the wavefront with a butterfly speed $v_B$~\cite{PhysRevLett.117.091602,Gu2017_1,Gu2017_2,PhysRevB.96.020406,PhysRevLett.121.024101,PhysRevB.100.045140,Mezei2020,10.21468/SciPostPhys.9.2.024}. However, when the Hamiltonian includes long-range (or power-law decaying) interactions proportional to $R^{-\alpha}$ with the distance $R$ between two particles, the wavefront can spread super-linearly with time~\cite{PhysRevX.3.031015,PhysRevLett.111.207202,PhysRevLett.111.260401,PhysRevLett.112.210601,PhysRevB.90.174204,PhysRevLett.119.170503,Cevolani_2016,Lepori_2017,PhysRevB.98.024302,PhysRevB.95.094205,PhysRevA.99.052332,PhysRevA.99.032114,PhysRevE.101.042118}. 
From the analogy of the short-range interacting systems,  
the following exponential growth of the OTOC may be inferred: 
\begin{align}
\label{exponential_growth_OTOC}
C(R,t) \propto e^{\lambda_L t}/R^\alpha.
\end{align}
It results in the so-called \textit{fast scrambling} which implies that local quantum information is spread over the entire regime of the system with a time scale of $t_s \approx \log(n)/\lambda_L$, where $n$ is the system size. 
Indeed, the well-known Lieb--Robinson bound~\cite{ref:Hastings2006-ExpDec,Nachtergaele2006} for long-range interacting systems gives the upper bound in the form of \eqref{exponential_growth_OTOC}.
Recent studies have focused on the universal laws of fast scrambling, specifically in the context of black hole physics~\cite{Sekino_2008,Lashkari2013,Bentsen6689}.
Starting with the exact solution of the Sachdev--Ye--Kitaev model~\cite{kitaev2014hidden,PhysRevD.94.106002}, 
 intensive studies have been conducted to determine the types of quantum many-body systems that permit/prohibit the fast scrambling~\cite{PhysRevB.94.035135,PhysRevB.95.134302,PhysRevB.98.134303,PhysRevLett.123.130601,harrow2019separation,PhysRevA.99.053620,PhysRevA.99.051803,chen2019operator,li2020fast,belyansky2020minimal,doi:10.1063/5.0022177}.

Fast scrambling implies that a system can relax arbitrarily fast under a local perturbation, 
whereas it is difficult to imagine that such extremely fast information propagation usually occurs in nature.
Systems with very large $\alpha$ are categorized as short-range systems, and hence, 
the OTOC cannot be described accurately for the entire regime of $\alpha$ by~\eqref{exponential_growth_OTOC}. Indeed, a more accurate description of the OTOC for long-range interacting systems may lead to the following polynomial growth~\cite{PhysRevB.98.134305,PhysRevB.100.064305,PhysRevA.99.010105,PhysRevLett.124.180601,PhysRevResearch.2.043047,PhysRevLett.114.157201,Matsuta2017,PhysRevA.101.022333,PhysRevX.9.031006,tran2019locality,chen2019finite,PhysRevX.10.031009,PhysRevX.10.031010} instead of an exponential growth~\eqref{exponential_growth_OTOC}:
\begin{align}
\label{polynomial_growth_OTOC}
C(R,t) \le \left( \lambda_L t/ R^\zeta \right)^{\tilde{\alpha}},
\end{align}
where $\zeta \le 1$ and $\zeta \tilde{\alpha}  \le \alpha$.
This inequality yields scrambling time that is algebraic with respect to the system size, i.e., $\approx n^{\zeta/D}/\lambda_L$.
For sufficiently large $\alpha$, several numerical~\cite{PhysRevB.98.134305,PhysRevB.100.064305,PhysRevA.99.010105,PhysRevLett.124.180601,PhysRevResearch.2.043047} and theoretical~\cite{PhysRevLett.114.157201,Matsuta2017,PhysRevA.101.022333,PhysRevX.9.031006,tran2019locality,chen2019finite,PhysRevX.10.031009,PhysRevX.10.031010} studies indicate polynomial growth.

From the above background, the following fundamental question naturally arises: \textit{what is the optimal condition for $\alpha$ to prohibit the fast scrambling of the OTOC given in \eqref{exponential_growth_OTOC}?}
Because numerical calculations have already indicated that polynomial growth of the OTOC might break down for $\alpha\le D$~\cite{PhysRevB.100.064305,PhysRevLett.124.180601}, we expect that the condition $\alpha>D$ is at least necessary. 
Moreover, this condition defines natural long-range interacting systems that are thermodynamically stable such that the total energy is extensive with regard to the system size and the thermodynamic limit is well-defined~\cite{dauxois2002dynamics,CAMPA200957}.

In previous studies, theoretical analyses have been mostly limited to the regime of $\alpha>2D$~\cite{PhysRevLett.114.157201,Matsuta2017,PhysRevA.101.022333,PhysRevX.9.031006,tran2019locality,chen2019finite,PhysRevX.10.031009,PhysRevX.10.031010}.
For $\alpha>2D$, Foss-Feig et al. proved that $\zeta$ in \eqref{polynomial_growth_OTOC} is lower-bounded by $\frac{\alpha-2D}{\alpha-D+1}$~\cite{PhysRevLett.114.157201}, which was improved to $\frac{\alpha-2D}{\alpha-D}$ in Refs.~\cite{PhysRevX.9.031006,tran2019locality}.
Furthermore, for $\alpha>2D+1$, even the existence of the finite butterfly speed (i.e., $\zeta=1$) has been proven in generic long-range interacting systems~\cite{chen2019finite,PhysRevX.10.031009,PhysRevX.10.031010}.
The sequence of these achievements has demonstrated that fast scrambling~\eqref{exponential_growth_OTOC} is prohibited in long-range interacting systems when $\alpha$ is above a threshold, i.e. $\alpha=2D$. 

In contrast, fast scrambling conditions in regimes of $D<\alpha\le2D$ are highly elusive.
In this regime, a sub-exponential speed of the quantum-state-transfer is in principle possible by a clever protocol employing quantum many-body long-range interactions~\cite{tran2020optimal}.
In addition, when exponent $\alpha$ approaches $D$, the effective system dimensions become infinitely large, 
and hence different physics can appear. 
For example, various studies on one-dimensional systems have shown that the long-range interactions can qualitatively change the fundamental physical properties for $\alpha \le2$ both in static~\cite{dyson1969,PhysRev.187.732,PhysRevLett.37.1577,PhysRevLett.87.137203,Kuwahara2020arealaw} and dynamical phases~\cite{PhysRevB.96.134427,PhysRevLett.120.130601}. 
Therefore, physics induced by long-range interactions in this regime is quite non-trivial and can yield unexpected consequences.
Nevertheless, various observations have indicated the prohibition of fast scrambling in this regime. 
As a partial solution, Tran et al. have disproved fast scrambling for a condition $\alpha>3/2$ in one dimension~\cite{PhysRevX.10.031009}.

In the present letter, we prove that under the condition $\alpha>D$ fast scrambling is prohibited in arbitrary long-range interacting systems.
Thus, by combining the counterexamples for $\alpha\le D$~\cite{PhysRevB.100.064305,PhysRevLett.124.180601}, we identify $\alpha>D$ as the optimal condition for the polynomial growth~\eqref{polynomial_growth_OTOC} of the OTOC (see also~\cite{Foot1}). 
As a general upper bound, we derive the polynomial growth of the OTOC with exponent $\zeta$ expressed as $\zeta=\frac{2\alpha-2D}{2\alpha-D+1}$. 
Our analyses consist of the following two parts: i) A simple connection technique for the unitary time operators for small times, which is utilized in Ref.~\cite{Kuwahara_2016_njp} and ii) the Lieb--Robinson bound for short-time evolution. Using these techniques, we can not only prove our main result, but also develop a considerably simple proof for the state-of-the-art Lieb--Robinson bound for $2D< \alpha\le 2D+1$ in~\cite{PhysRevX.9.031006,tran2019locality}. 
Our result verifies the empirical hypothesis that thermodynamically natural class of long-range interactions cannot induce fast scrambling.

\begin{figure}[]
\centering
{
\includegraphics[clip, scale=0.38]{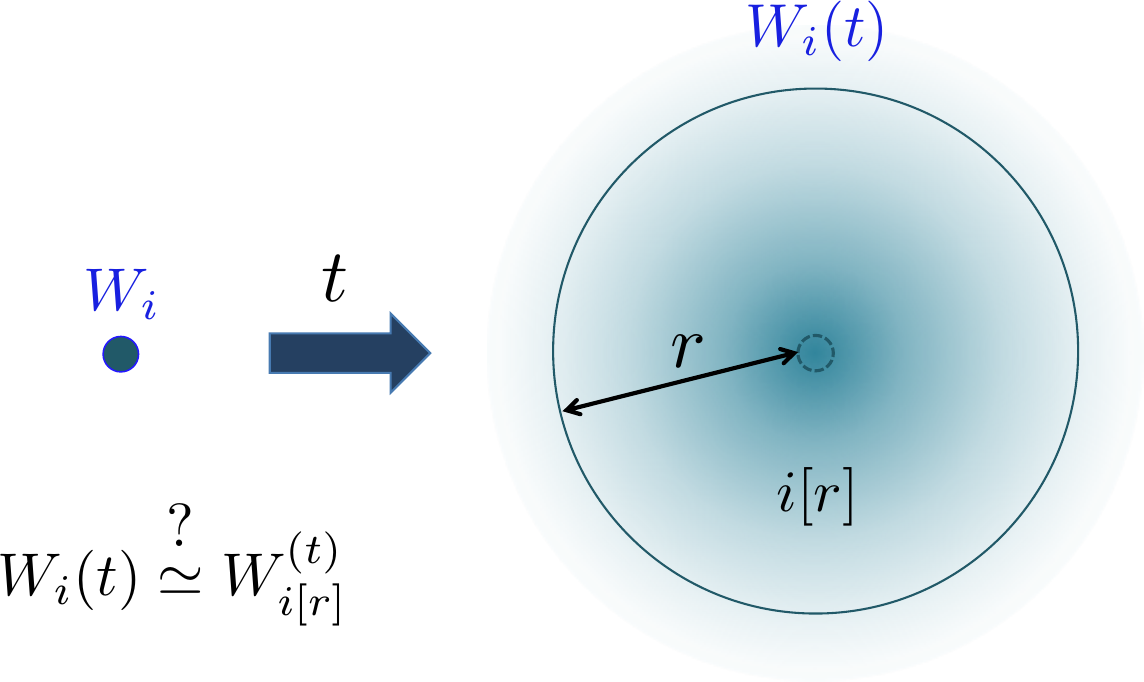}
}
\caption{(color online) The OTOC~\eqref{OTOC_def} roughly determines the spreading of local operator $W_i$ by time evolution. We aim to approximate $W_i(t)$ in a local region $\bal{i}{r}$, which has a maximum distance of $r$ from site $i$ [Eq.~\eqref{def:bal_X_r}]: If 
operator $W_i(t)$ is well approximated by using $W_{i[r]}^{(t)}$ as long as 
$t\lesssim \orderof{r^{\zeta}}$ ($\zeta<1$), 
the OTOC exhibits polynomial growth, as in Eq.~\eqref{polynomial_growth_OTOC}, because of \eqref{OTOC_C_R_t}.
}
\label{fig:local_approx}
\end{figure}

\sectionprl{Setup and main result}
Let us consider a quantum spin system with $n$ spins, where each spin is located on one vertex of the $D$-dimensional graph (or $D$-dimensional lattice) with $\Lambda$ of the total spin set, i.e., $|\Lambda| = n$.
For simplicity, we consider (1/2)-spin systems; however, the extension to a general finite spin dimension $d$ is straightforward. For a partial set $X\subseteq \Lambda$, we denote the cardinality, i.e., the number of vertices contained in $X$, by $|X|$ (e.g., $X = \{i_1,i_2,\ldots, i_{|X|}\}$). Further, we denote the complementary subset of $X$ as $X^\co :=  \Lambda\setminus X$.
For two arbitrary spins $i$ and $i'$, we define distance $\dist_{i,i'}$ as the shortest path length on the lattice that connects $i$ and $i'$.
We define $i[r]$ as the ball region with radius $r$ from site $i$ (Fig.~\ref{fig:local_approx}). 
\begin{align}
\bal{i}{r}:= \{i'\in \Lambda| \dist_{i,i'} \le r \} , \label{def:bal_X_r}
\end{align}
where $\bal{i}{0} = i$ and $r$ is an arbitrary positive integer.

We consider a general system having at most $k$-body long-range interactions with finite $k$. 
For example, we give the Hamiltonian with $k=2$, which is described as
\begin{align}
H=\sum_{i<i'}  h_{i,i'} +  \sum_{i=1}^n h_i ,\quad \|h_{i,i'}\|\le \frac{J_0}{(\dist_{i,i'}+1)^\alpha} \label{Ham:general_power}
\end{align} 
for $\forall i,i' \in \Lambda$, where $\{h_{i,i'}\}_{i<i'}$ are interaction operators acting on the spins $\{i,i'\}$, and $\|\cdots\|$ is the operator norm.
One of the simple examples is the long-range transverse Ising model, which has a form of Eq.~\eqref{Ham:general_power} by choosing $h_{i,i'} =  J \sigma_i^x  \sigma_{i'}^x /\dist_{i,i'}^\alpha $ and $h_i=B\sigma_i^z$. 
Such long-range interactions have been realized in various experimental setups such as atomic, molecular, and optical systems~\cite{
bendkowsky2009observation,
RevModPhys.80.885,
RevModPhys.82.2313,
yan2013observation,
PhysRevLett.108.210401,
britton2012engineered,
Islam583,
zeiher2016many,
PhysRevX.7.041063,
bernien2017probing,
zhang2017observation,
Neyenhuise1700672,
PhysRevLett.122.150601,
tan2019observation}.
In this letter, we are in particular interested in the regime of $D<\alpha\le 2D$, which is also experimentally important as it includes several realistic long-range interactions, such as dipole--dipole interactions ($D=2$, $\alpha=3$) and van der Waals interactions ($D=3$, $\alpha=6$).

%

In our analyses, we focus on time evolution by the Hamiltonian $H$.
A key strategy for estimating the OTOC is using the local approximation of the time-evolved operator $W_i(t):=e^{iHt} W_i e^{-iHt}$ (Fig.~\ref{fig:local_approx}). 
We approximate the operator $W_i(t)$ using another operator $W_{i[r]}^{(t)}$ which is supported on the local subset $i[r]$. The error of this approximation is estimated by
 \begin{align}
 \label{local approximation}
\normb { W_i(t)- W_{i[r]}^{(t)}  }_p , 
\end{align}
where $\|\cdots\|_p$ is the Schatten-$p$ norm, which is defined as $\|O\|_p:= [\tr (O^\dagger O)^{p/2}]^{1/p}$. 
For $p = \infty$, the Schatten norm $\|\cdots\|_\infty$ corresponds to the standard operator norm, while the case of $p = 2$ corresponds to the Frobenius norm, which is of interest. 
For an arbitrary operator $V_{i'}$ with $\dist_{i,i'} = R$, one can easily show
\begin{align}
\label{OTOC_C_R_t}
C(R,t)\le  4 \normb { W_i(t)- W_{i[R-1]}^{(t)}  }_F^2, 
\end{align}
where we define the normalized Frobenius norm $\|\cdots\|_F :=  \|\cdots\|_2/ [\tr (\hat{1})]^{1/2}$ and use $[W_{i[R-1]}^{(t)},V_{i'}] = 0$ for $\dist_{i,i'} = R$.

Our main result provides the efficiency guarantee for the local approximation of a time-evolved operator $W_i(t)$ in the region $\bal{i}{r}$ (see~\cite[Section S.II]{Supplement_long_range_LR} for more details).
\begin{theorem}  \label{thm_main_poly_light_OTOC}
Let us consider Hamiltonians with few-body interactions and power-law decay exponent $\alpha >D$. 
Then, for an arbitrary operator $W_i$ ($\|W_i\| = 1$) and the corresponding time evolution of $W_i(t)$, there exists an operator $W^{(t)}_{i[r]}$ that approximates $W_i(t)$ on a region $i[r]$ as
\begin{align}
 \label{ineq:corol_main_poly_light}
\norm{ W_i(t) - W^{(t)}_{i[r]} }_F \le C r^{-\alpha+D} t^{\alpha-\frac{D-1}{2}} ,
\end{align} 
where $C$ is an $\orderof{1}$ constant.
\end{theorem}
\noindent 
From the inequalities in \eqref{OTOC_C_R_t} and \eqref{ineq:corol_main_poly_light}, we obtain the upper bound of the OTOC as 
\begin{align}
C(R,t)\lesssim  \left( \frac{C' t}{R^{\frac{2\alpha-2D}{2\alpha-D+1}}} \right)^{\alpha-\frac{D-1}{2}},  \notag 
\end{align}
where $C'$ is a constant of $\orderof{1}$. 
This gives the polynomial growth in \eqref{polynomial_growth_OTOC} with $\zeta=\frac{2\alpha-2D}{2\alpha-D+1}$ and $\tilde{\alpha}=\alpha-(D-1)/2$.

In the above theorem, we consider an on-site operator $W_i$; however, the theorem can be generalized to an operator $W_X$ supported on an arbitrary subset $X \subset \Lambda$. 
Let us consider the case where the subset $X$ satisfies $X \subseteq i[r_0]$ for particular choices of $i$ and $r_0$. Then, for $W_X(t)$, we obtain an inequality that is similar to~\eqref{ineq:corol_main_poly_light} as 
\begin{align}
\norm{ W_X(t) - W^{(t)}_{i[r_0+r]} }_F \le \frac{C t^{\alpha-\frac{D-1}{2}} (r+r_0)^{\frac{D-1}{2}}}{r^{\alpha-\frac{D+1}{2}}}.\notag 
\end{align} 
For $D = 1$, the above inequality reduces to 
\begin{align}
\norm{ W_X(t) -W^{(t)}_{i[r_0+r]} }_F \le  \frac{C t^{\alpha}}{r^{\alpha-1}}. \notag 
\end{align}

\begin{figure}[]
\centering
{
\includegraphics[clip, scale=0.33]{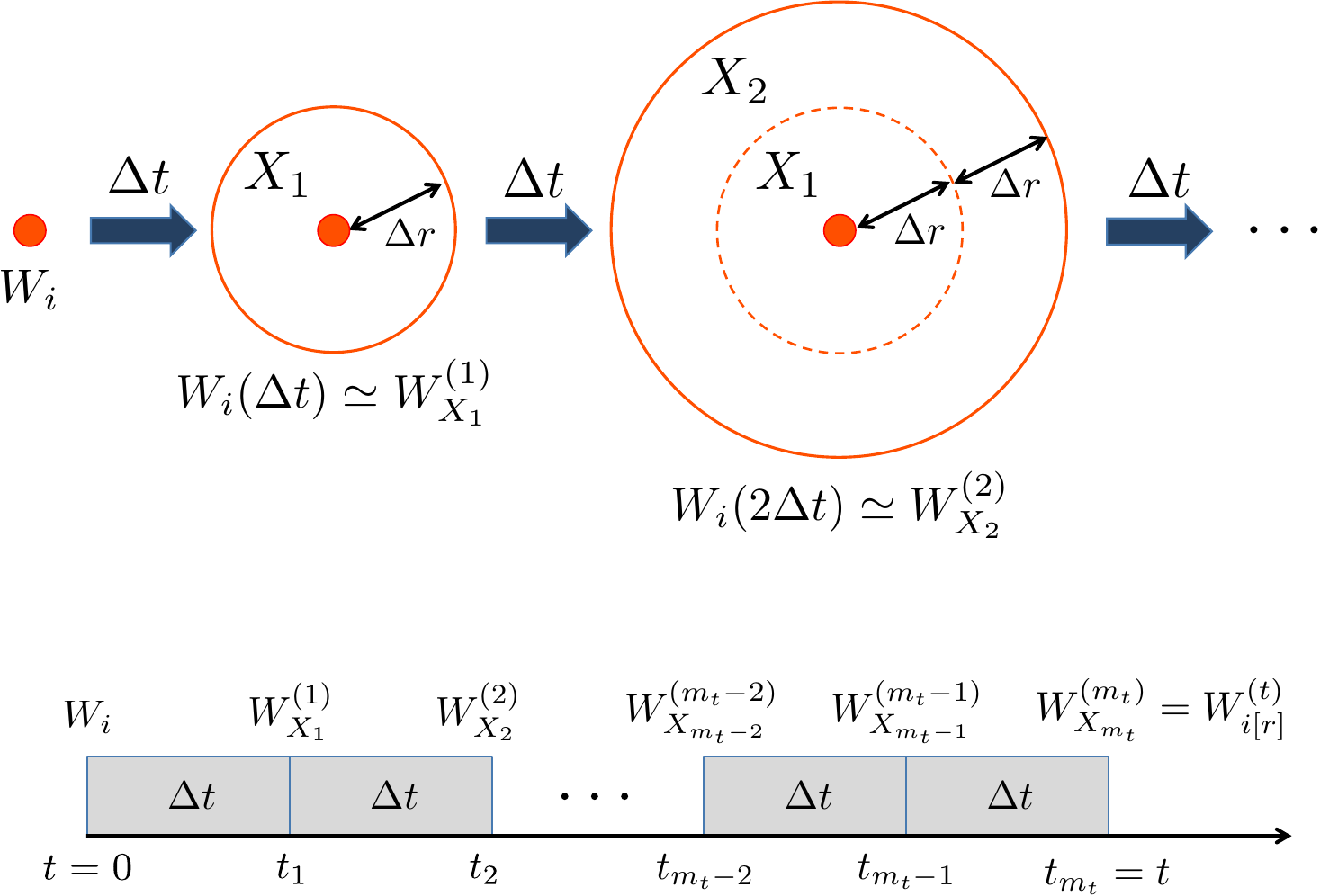}
}
\caption{(color online) We decompose time $t$ and length $r$ to $m_t$ pieces, namely $\Delta t := t/m_t$ and $\Delta r:= r/m_t$. 
We start from time evolution $W_i(\Delta t)$ and approximate it by $W_{X_1}^{(1)}$, which is supported on an extended region $X_1$ as in Eq.~\eqref{Choice_Delta_r_X_m}. Then, we iteratively approximate $W_{X_m}^{(m)} (\Delta t)$ by $W_{X_{m+1}}^{(m+1)}$, which finally yields the  approximation~\eqref{unitary_connect_upper_bound}. 
The main advantage of this method is that we need to estimate the local approximation of the time-evolved operators only for a short time. 
 }
\label{fig_unitary_connect}
\end{figure}

\sectionprl{Concept of the proof}
A central technique in our proof is the connection of unitary time evolutions addressed in Ref.~\cite{Kuwahara_2016_njp} (Fig.~\ref{fig_unitary_connect}). Following reference ~\cite{Kuwahara_2016_njp}, we decompose the time to $m_t$ pieces, and we define $t_m := m \Delta t$  and $t_{m_t} := t$ where $\Delta t = t/m_t$.
We assume $\Delta t$ as a small constant. For fixed $r$ and $i\in \Lambda$, we define lengths $\Delta r$, $r_m$, and subset $X_m$ as 
\begin{align}
\label{Choice_Delta_r_X_m}
\Delta r :=r/m_t   ,  \quad X_m:= i[m\Delta r]. 
\end{align}
Using these notations, we approximate $W_i(t_m)$ with another operator supported on subset $X_m$. 

For the approximation, we adopt the following recursive procedure.
For $m = 1$, we define operator $W_{X_1}^{(1)}$ as an approximation of $W_i(\Delta t)$ onto the subset $X_1$:
\begin{align}
W_{X_1}^{(1)} := W_i(\Delta t,X_1) , \notag 
\end{align}
where we define notation $W_i(t,X_1)$ as 
 \begin{align}
\label{def_W_i_t_X_1}
W_i(t,X_1):= \frac{1}{\tr_{X_1^\co}(\hat{1})} \tr_{X_1^\co} \left[W_i(t)\right] \otimes \hat{1}_{X_1^\co}.
\end{align}
Note that $W_i(\Delta t,X_1)$ is now supported on subset $X_1$.
For $m = 2$, we adopt the second-step approximation 
$W_{X_2}^{(2)} := W_{X_1}^{(1)}(\Delta t, X_2),$ 
which is similar to \eqref{def_W_i_t_X_1}. We then obtain the approximation error as
\begin{align}
\label{approx_second_m=2}
&\bigl \| W_i(2\Delta t) - W_{X_2}^{(2)} \bigr\|_p \notag \\
& \le  \bigl \|  W_i(2\Delta t) -W_{X_1}^{(1)}(\Delta t) + W_{X_1}^{(1)}(\Delta t)  - W_{X_2}^{(2)} \bigr \|_p \notag \\
&\le \bigl \|  W_i(\Delta t) -W_{X_1}^{(1)}\bigr \|_p + \bigl\|W_{X_1}^{(1)}(\Delta t)  - W_{X_2}^{(2)} \bigr \|_p, 
\end{align}
with $W_{X_1}^{(1)}(\Delta t):= e^{i H\Delta t} W_{X_1}^{(1)} e^{-i H\Delta t }$, where we use the triangle inequality and unitary invariance for the Schatten-$p$ norm. 

By repeating this procedure, we define operator $W_{X_m}^{(m)}$ recursively as $W_{X_{m}}^{(m)}= W_{X_{m-1}}^{(m-1)} (\Delta t,X_m)$. Then, similar to ~\eqref{approx_second_m=2}, we obtain the following inequality: 
 \begin{align}
 \label{unitary_connect_upper_bound} 
&\normb{W_i(m_t\Delta t) - W_{X_{m_t}}^{(m_t)} }_p \notag \\
& \le  \sum_{m=0}^{m_t-1} \normb{ W_{X_m}^{(m)} (\Delta t) - W_{X_{m+1}}^{(m+1)}  }_p,
\end{align}
where we define $W_{X_0}^{(0)}:= W_i$.
The problem now reduces to the estimating the approximation error of $\| W_{X_m}^{(m)} (\Delta t)-W_{X_{m+1}}^{(m+1)}\|_p$ only for short-time evolution, which is a critical point to derive our main results.

As the simplest exercise, let us consider the case with $p = \infty$, which provides the standard operator norm. The resulting wavefront shape for information propagation is the same as that obtained in~\cite{PhysRevX.9.031006,tran2019locality}; however, our derivation is considerably simpler and can be applied to a more general class of Hamiltonians. 
For the short-time evolution, we can utilize the well-known simple Lieb--Robinson bound as in \cite{ref:Hastings2006-ExpDec,Nachtergaele2006}. 
Using their results, we can readily derive the following approximation error (see \cite[Section S.III A]{Supplement_long_range_LR} for the derivation):
\begin{align}
\label{Hastings_Koma_bound}
\normb{  W_{X_m}^{(m)} (\Delta t) - W_{X_{m+1}}^{(m+1)}  }_\infty \le
c   |\partial X_m| e^{c'\Delta t} (\Delta r)^{-\alpha+D+1}, 
\end{align}
where $c$ and $c'$ are the constants of $\orderof{1}$, which depend on only the details of the system. Note that $\partial X_m$ is the surface region of subset $X_m$.  
For a sufficiently large $\Delta t$, the bound~\eqref{Hastings_Koma_bound} eventually yields an exponential growth; however, $\Delta t$ is now selected to be as small as $\orderof{1}$, and hence, $e^{c'\Delta t}$ is given by a constant.

Thus, by introducing geometric parameter $\gamma$ that yields $|\partial X_m| \le |\partial i[r]|\le  \gamma r^{D-1}$, we obtain 
\begin{align}
\normb{W_{X_m}^{(m)} (\Delta t)-W_{X_{m+1}}^{(m+1)}}_\infty \le  \tilde{c} r^{2D-\alpha}t^{\alpha-D-1}, \notag
\end{align}
where $\tilde{c}:=c\gamma e^{c'\Delta t} (\Delta t)^{-\alpha+D+1}$, and we use $\Delta r= \Delta t(r/t)$. Therefore, we reduce the upper bound in~\eqref{unitary_connect_upper_bound} to
 \begin{align}
 \label{derivation_previous_Lieb_Robinson}
\normb{W_i(t) - W_{i[r]}^{(m_t)} }_\infty& \le  \tilde{c}'  r^{2D-\alpha} t^{\alpha-D} ,
\end{align}
where $\tilde{c}':=\tilde{c}/\Delta t$, and we use $m_t=t/\Delta t$.  
The time step, $\Delta t$, is selected as an $\orderof{1}$ constant, and hence, $\tilde{c}'$ is also an $\orderof{1}$ constant. 
Using the upper bound, information propagation is restricted to a region with diameter $R\approx |t|^{\frac{\alpha-D}{\alpha-2D}}$, which is the same as the state-of-the-art estimation obtained in~\cite{PhysRevX.9.031006,tran2019locality}, namely the improved version of~\cite{PhysRevLett.114.157201,Matsuta2017,PhysRevA.101.022333}.
Note that the result above is more general; we \textit{do not} have to assume the few-body interactions of the Hamiltonian in deriving \eqref{Hastings_Koma_bound} because the upper bound in~\eqref{Hastings_Koma_bound} is applied to the Hamiltonians without the assumption of few-body interactions  (see \cite[Assumption~2.1]{ref:Hastings2006-ExpDec}).

Finally, we explain why the condition of $\alpha>2D$ appears instead of $\alpha>D$ to obtain a meaningful upper bound.   
This condition originated from coefficient $|\partial X_m|$ in \eqref{Hastings_Koma_bound}. When we consider the time evolution of an operator supported on subset $X \subset \Lambda$ (e.g., $O_X$), the Lieb--Robinson bound unavoidably includes the subset dependence~\cite{ref:LR-bound72,PhysRevLett.97.050401,ref:Nachtergaele2006-LR}. 
This subset dependence is the primary obstacle that resists the rigorous proof of the polynomial growth of the information propagation for $\alpha<2D$. In the case where the Frobenius norm ($p = 2$) is considered, this subset dependence is significantly improved, as shown in~\eqref{Ineq:thm_local_approximation_frobenius}. This provides a breakthrough in deriving the strictest condition, namely $\alpha>D$, for the polynomial growth of the OTOC.



\sectionprl{Proof of Theorem~\ref{thm_main_poly_light_OTOC} (Case with $p = 2$ and $\alpha>D$)}
For proving our main theorem, we start from the inequality in~\eqref{unitary_connect_upper_bound}. Thus, our task is to derive a local approximation for short-time evolution. Here, let $O_X$ be an arbitrary operator on subset $X$ with $\|O_X\|=1$. We aim to approximate $O_X(t)$ by $O_X(t, X[r])$, where $X[r]$ is an extended subset defined as $X[r]:=\bigcup_{i\in X} i[r]$. The key technical ingredient is the following inequality for short-time evolution in terms of the Frobenius norm~(see \cite[Theorem~3]{Supplement_long_range_LR})
\begin{align}
\label{Ineq:thm_local_approximation_frobenius}
&\| O_X(t)- O_X(t, X[r]) \|_F  \notag \\
&\le c_0 |t|  \sqrt{|\partial X[r] | \cdot r^{-2\alpha+D+1}},
\end{align}
with $c_0$ as an $\orderof{1}$ constant, where $\partial X[r]$ is the surface region of $X[r]$, and time $t$ is assumed to be smaller than a certain threshold. 
Most parts of the proof are dedicated to deriving~\eqref{Ineq:thm_local_approximation_frobenius}, as shown in the Supplementary Material (\cite[Sections S.IV and S.V]{Supplement_long_range_LR}).

With the inequality in~\eqref{Ineq:thm_local_approximation_frobenius}, we can easily prove the main theorem~\ref{thm_main_poly_light_OTOC} in the same manner as that used for deriving \eqref{derivation_previous_Lieb_Robinson} for $p = \infty$. 
Here, $\Delta t$ is sufficiently small such that the inequality~\eqref{Ineq:thm_local_approximation_frobenius} holds.  
Applying inequality~\eqref{Ineq:thm_local_approximation_frobenius} to \eqref{unitary_connect_upper_bound}, we obtain
\begin{align}
\normb{W_{X_m}^{(m)} (\Delta t)-W_{X_{m+1}}^{(m+1)}}_F  \le  \tilde{c}_0 r^{-\alpha+D}t^{\alpha-\frac{D+1}{2}}\notag 
\end{align}
with $\tilde{c}_0$ being an $\orderof{1}$ constant, where we use $W_{X_{m+1}}^{(m+1)}= W_{X_m}^{(m)} (\Delta t,X_m[\Delta r])$ and $|\partial (X_m[\Delta r])| \le |\partial (i[2r])| \le \gamma (2r)^{D-1}$.
The above inequality reduces inequality in~\eqref{Ineq:thm_local_approximation_frobenius} to the main inequality given in~\eqref{ineq:corol_main_poly_light} using $m_t:= t/\Delta t$ as 
\begin{align}
\normb{W_i(t) - W_{i[r]}^{(m_t)} }_F& \le  (\tilde{c}_0/\Delta t)  r^{-\alpha+D} t^{\alpha-\frac{D-1}{2}}. \notag 
\end{align}
This completes the proof of Theorem~\ref{thm_main_poly_light_OTOC}. $\square$

\sectionprl{Conclusion}
In this work, we investigated the polynomial growth of the OTOC represented in~\eqref{polynomial_growth_OTOC} for all long-range interacting systems with $\alpha>D$, where the existence of a well-defined thermodynamic limit is ensured.
We comprehensively disproved fast scrambling in this natural class of long-range interactions. 
Our results indicate the lower bound of the scrambling time as $n^{\zeta/D}$ with $\zeta=\frac{2\alpha-2D}{2\alpha-D+1}$.

This study has two future directions. 
First, our condition of $\alpha>D$ for the polynomial growth of the OTOC is expected to be qualitatively tight; however, the quantitative estimation of $\zeta$ still has scope for improvement. In particular, it is an intriguing problem to identify the critical value of $\alpha_c$ above which the ballistic propagation of information scrambling (i.e., $\zeta=1$) is ensured. For the operator norm [i.e., $p=\infty$ in Eq.~\eqref{local approximation}], the critical $\alpha_c$ is proven to be equal to $2D+1$~\cite{chen2019finite,PhysRevX.10.031009,PhysRevX.10.031010}.
For the Frobenius norm, it has been conjectured that the critical $\alpha_c$ is equal to $3D/2+1$, where the case of $D = 1$ has been indeed proved~\cite{PhysRevX.10.031009}. We hope that our current analysis will be further refined to identify the optimal value of $\zeta$ in the future.

Second, we considered the most common form of the OTOC in~\eqref{OTOC_def}, which adopts the average for a uniformly mixed state. 
In experimental application, if we would be able to prepare the uniform mixed state as the initial state, 
Theorem~\ref{thm_main_poly_light_OTOC} appropriately predicts the growth of the OTOC.
On the other hand, if the initial state is prepared as a finite temperature result, 
we need to consider the following generalization for a finite-temperature state:
\begin{align}
C _\beta(x,t) :=\frac{1}{\tr (e^{-\beta H})} \tr (e^{-\beta H} [W_i(t),V_{i'}]^\dagger [W_i(t),V_{i'}] ) \notag .
\end{align}
The inequality in~\eqref{unitary_connect_upper_bound} is applied to this case, and we expect that the same polynomial growth can be obtained above a temperature threshold by using the cluster expansion technique~\cite{PhysRevX.4.031019,PhysRevLett.124.220601}. 

Finally, throughout the paper, we consider the Hamiltonian dynamics $e^{-iHt}$.
It is an intriguing to extend our result to Markovian quantum dynamics~\cite{PhysRevLett.104.190401,PhysRevLett.108.230504}.
If the uniform mixed state is a steady state, our formalism in~\eqref{unitary_connect_upper_bound} is applied and we expect to derive a similar upper bound for the OTOC.

\begin{acknowledgments}
The work of T. K. was supported by the RIKEN Center for AIP and JSPS KAKENHI (Grant No. 18K13475). 
TK gives thanks to God for his wisdom.
K.S. was supported by JSPS Grants-in-Aid for Scientific Research (JP16H02211 and JP19H05603).
\end{acknowledgments}

\bibliography{Long_range_LR}

\begin{thebibliography}{116}
\bibitem [{Foo()}]{Foot1}%
  \BibitemOpen
  \href@noop {} {}Our condition of $\alpha>D$ is applied to general quantum
  many-body systems to satisfy the polynomial growth of the
  OTOC~\eqref{polynomial_growth_OTOC}. However, this does not means that for
  $\alpha<D$ the systems necessarily show fast scrambling. Indeed, there exists
  a class of long-range interacting
  systems~\cite{PhysRevA.102.022402,PhysRevA.102.010401} which show the
  polynomial growth even for $\alpha<D$.\BibitemShut {Stop}%
\end{thebibliography}

\renewcommand\thefootnote{*\arabic{footnote}}

\clearpage
\newpage

\addtocounter{section}{0}

\addtocounter{equation}{-15}

\renewcommand{\theequation}{S.\arabic{equation}}

\renewcommand{\thesection}{S.\Roman{section}}
\begin{widetext}

\begin{center}
{\large \bf Supplementary Material for ``Absence of fast scrambling in thermodynamically stable long-range interacting systems''}\\
\vspace*{0.3cm}
Tomotaka Kuwahara$^{1,2}$, Keiji Saito$^{3}$ \\
\vspace*{0.1cm}
$^{1}${\small \it Mathematical Science Team, RIKEN Center for Advanced Intelligence Project (AIP),1-4-1 Nihonbashi, Chuo-ku, Tokyo 103-0027, Japan}\\
$^{2}${\small \it Interdisciplinary Theoretical \& Mathematical Sciences Program (iTHEMS) RIKEN 2-1, Hirosawa, Wako, Saitama 351-0198, Japan} \\
$^{3}${\small \it Department of Physics, Keio University, Yokohama 223-8522, Japan} 
\end{center}

\tableofcontents

\section{Set up and Preliminaries}

\subsection{Notations}

We here recall the setup.
We consider a quantum spin system with $n$ spins, where each of the spin sits on a vertex of the $D$-dimensional graph (or $D$-dimensional lattice) with $\Lambda$ the total spin set, namely $|\Lambda|=n$.
For the simplicity, we consider (1/2)-spin systems, but the extension to a general finite spin dimension $d$ is straightforward; we only let $n\to n\log(d)$ and $k\to k\log(d)$, where $k$ will be defined in Eq.~\eqref{supp_def:Ham}.
For a partial set $X\subseteq \Lambda$, we denote the cardinality, that is, the number of vertices contained in $X$, by $|X|$ (e.g. $X=\{i_1,i_2,\ldots, i_{|X|}\}$).
We also denote the complementary subset of $X$ by $X^\co := \Lambda\setminus X$.

 \begin{figure}[tt]
\centering
\includegraphics[clip, scale=0.5]{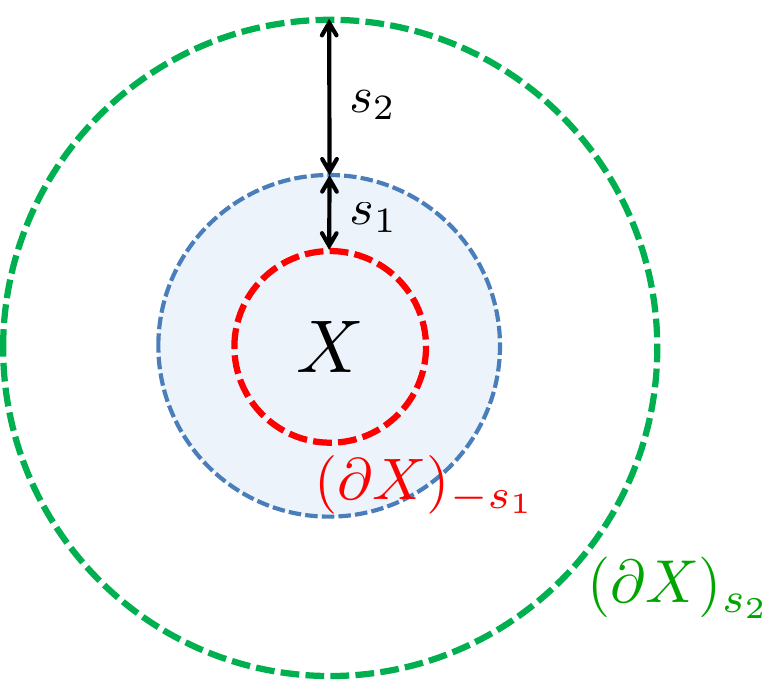}
\caption{Schematic picture of the definition of $(\partial X)_s$ for positive and negative $s$. 
}
\label{supp_fig_partial_X_s}
\end{figure}

For arbitrary subsets $X, Y \subseteq \Lambda$, we define $\dist_{X,Y}$ as the shortest path length on the graph that connects $X$ and $Y$; that is, if $X\cap Y \neq \emptyset$, $\dist_{X,Y}=0$. 
When $X$ is composed of only one element (i.e., $X=\{i\}$), we denote $\dist_{\{i\},Y}$ by $\dist_{i,Y}$ for the simplicity.
We also define $\diam(X)$ as follows: 
\begin{align}
\diam(X):  =\max_{i,i'\in X} (\dist_{i,i'}).
\end{align}
For an arbitrary subset $X\subset \Lambda$, we denote the surface region of $X$ by $\partial X$. 
Moreover, we define $(\partial X)_s$ as follows (see Fig.~\ref{supp_fig_partial_X_s}): 
\begin{align}
\label{supp_notation_partial_X_s}
(\partial X)_s := 
\begin{cases}
\{i\in X| \dist_{i,\partial X} =s\} &\for s\le 0, \\
\{i\in X^\co | \dist_{i,\partial X} =s\} &\for s> 0, 
\end{cases}
\end{align}
where $(\partial X)_0= \partial X$ and we have 
\begin{align}
X = \bigcup_{s=-\infty}^0 (\partial X)_{s} ,\quad \Lambda= \bigcup_{s=-\infty}^\infty  (\partial X)_s .
\end{align}

For a subset $X\subseteq \Lambda$, we define the extended subset $\bal{X}{r}$ as
\begin{align}
\bal{X}{r}:= \{i\in \Lambda| \dist_{X,i} \le r \} = \bigcup_{s=-\infty}^r (\partial X)_{s}, \label{supp_def:bal_X_r}
\end{align}
where $\bal{X}{0}=X$ and $r$ is an arbitrary positive number (i.e., $r\in \mathbb{R}^+$).

We introduce a geometric parameter $\gamma$ which is determined only by the lattice structure.
We define $\gamma \ge 1$ as a lattice constant which gives for $X=i[r]$ ($r\ge1$)
\begin{align}
\label{supp_parameter_gamma_X}
|X| \le \gamma r^D ,\quad |\partial X| \le \gamma r^{D-1}.
\end{align}
By using the constant $\gamma$, we can derive the following inequality which we frequently use in the analyses:
\begin{align}
\label{supp_ineq_sum_decay_func}
\sum_{i\in \Lambda: \dist_{i,i'}>r} (\dist_{i,i'}+1)^{-a} &= \sum_{s=r+1}^\infty \sum_{i\in \Lambda: \dist_{i,i'}=s}(s+1)^{-a} \notag \\
&\le \gamma \sum_{s=r+1}^\infty (s+1)^{-a+D-1} \le \gamma \int_{r+1}^\infty x^{-a+D-1} dx = \frac{\gamma}{a-D} (r+1)^{-a+D}
\end{align}
for a fixed $i' \in \Lambda$ and $a>D$, where we use $\{i\in \Lambda: \dist_{i,i'}=s\} = \partial i'[s]$. 

\subsection{Long-range Hamiltonians}

We consider a $k$-local Hamiltonian as 
\begin{align}
H= \sum_{|Z| \le k} h_Z, \label{supp_def:Ham}
\end{align}
where each of the interaction terms $\{h_Z\}_{|Z|\le k}$ acts on the spins on $Z \subset \Lambda$.
In previous studies such as Refs~\cite{ref:Hastings2006-ExpDec,PhysRevX.10.031010} or the inequality which will be derived in~\eqref{supp_derivation_previous_Lieb_Robinson}, we do not need the assumption of the $k$-locality (i.e., $|Z|=\orderof{1}$), but
in the proof of our main results (i.e., Theorem~\ref{supp_thm_main_poly_light}), the $k$-locality plays a crucial role.
We do not explicitly consider the time-dependence of the Hamiltonians, but all the analyses can be generalized to the time-dependent Hamiltonians.

In order to characterize the long-range interaction of the Hamiltonian, we impose the following assumption for the Hamiltonian:
\begin{assump}[Power-law decaying interactions]  \label{supp_lem:power_law_k_local}
We assume the power-law decay of the interaction in the following senses:
\begin{align}
&\sup_{i,i' \in \Lambda: \dist_{i,i'}=r}\sum_{Z: Z\supset \{i,i'\}} \| h_Z\| \le  J_0 (r+1)^{-\alpha}  \label{supp_alternative_basic_assump_power}
\end{align}
with
\begin{align}
\alpha>D,
\end{align}
where $\|\cdots\|$ denotes the operator norm and the parameter $J_0$ is an $\orderof{1}$ constant which does not depend on the system size $n$.
Here, $\sum_{Z:Z\supset \{i,i'\}}$ means the summation which picks up all the subsets $Z\subset \Lambda$ which include $\{i,i'\}$
\end{assump}
\noindent

We define the parameter $\tilde{g}$ which we often use:
\begin{align}
\label{supp_tilde_g__def}
\tilde{g}:=\max(gk, \lambda J) , 
\end{align}
where $J:=3^{k/2}J_0$, $g$ is a one-site energy which is defined in \eqref{supp_g_extensive} and \eqref{supp_g_extensive_explicit}, and $\lambda$ is defined as an $\orderof{1}$ constant satifying 
\begin{align}
\sum_{i_0 \in \Lambda}  (\dist_{i,i_0}+1)^{-\alpha}  (\dist_{i_0,i'}+1)^{-\alpha}  \le \lambda (\dist_{i,i'}+1)^{-\alpha}   ,
\end{align}
for all the pairs $\{i,i'\}\subset \Lambda$.

\subsection{Generalized H\"older inequality for Schatten norm} 

For an arbitrary operator $O$, we define the Schatten-$p$ norm as follows: 
\begin{align}
\|O\|_p := \left[\tr (O^\dagger O)^{p/2} \right]^{1/p}. \label{supp_sup_def:Schatten p norm}
\end{align}
Note that $\|O\|_1$ corresponds to the trace norm and $\|O\|_\infty$ corresponds to the standard operator norm (i.e., the maximum singular value of $O$).
We often denote $\|O\|_\infty$ by $\|O\|$ for simplicity.
Especially, for $p=2$, the Schatten-2 norm corresponds to the Frobenius norm, namely $\|O\|_2=\sqrt{\tr(O^\dagger O)}$.  
Throughout the analyses, we utilize the notation $\|\cdots\|_F$ as the following normalized Frobenius norm:
\begin{align}
\label{supp_normalized_frobenius_norm}
\|O\|_F := \sqrt{ \tilde{\tr} (O^\dagger O)} = \sqrt{\frac{\tr (O^\dagger O)}{\tr(\hat{1}) }}  ,
\end{align}
where we define $\tilde{\tr}(\cdots)$ as $\tr(\cdots)/\tr(\hat{1})$.

For a general Schatten $p$ norm, we can obtain the following generalized H\"older inequality (see, for example Ref.~\cite[Proposition 2.5]{sutter2018approximate}):
\begin{align}
\left \| \prod_{j=1}^s O_j \right \|_p \le \prod_{j=1}^s \|O_j \|_{p_j} , \label{supp_sup_generalized_Holder_ineq}
\end{align}
where $\sum_{j=1}^s 1/p_j =1/p$.
From the inequality, we can immediately obtain 
\begin{align}
\left \| O_1 O_2 \right \|_F \le \|O_1\|_F \|O_2\|,
 \label{supp_sup_generalized_Holder_ineq_operator_norm}
\end{align}
where we set $p_1=2$ and $p_2=\infty$ in \eqref{supp_sup_generalized_Holder_ineq}.

\subsection{Local approximation of time-evolved operators}

We consider an operator $W_X$ which is defined on a subset $X$. 
For the time-evolved operator $W_X(t)$, we define 
$W_X(t,\tilde{X})$ as the local approximation of $W_X(t)$ onto the subset $\tilde{X}$:
\begin{align}
W_X(t,\tilde{X})  := \frac{1}{\tr_{\tilde{X}^\co}(\hat{1})} \tr_{\tilde{X}^\co} \left[W_X(t)\right] \otimes \hat{1}_{\tilde{X}^\co},
\label{supp_def:W_X_local_approx}
\end{align}
where $\tr_{\tilde{X}^\co}(\cdots)$ is the partial trace with respect to the subset $\tilde{X}^\co$.
The definition implies that the operator $W_X(t,\tilde{X})$ is supported on the subset $\tilde{X}$ and it also satisfies $\|W_X(t,\tilde{X})\|\le \|W_X\|$.
In our paper, we aim to estimate the approximation error between $W_X(t)$ and $W_X(t, X[r])$ for the Schatten-$p$ norm.
As for the operator norm (i.e., $\|\cdots\|_p$ for $p=\infty$), the upper bound has been given by Bravyi, Hastings and Verstaete as follows~\cite{PhysRevLett.97.050401}:
\begin{align}
\label{supp_BHV_bound}
\| W_X(t)- W_X(t, X[r]) \| \le \sup_{U_{X[r]^\co}} \|[W_X(t), U_{X[r]^\co}] \|,
\end{align}
where $\sup_{U_{X[r]^\co}}$ is taken from all the unitary operators on the subset $X[r]^\co$.

\section{Main results}

We here show our main theorem:
\begin{theorem}   \label{supp_thm_main_poly_light}
Let $\|\cdots \|_F$ as the normalized Frobenius norm defined in Eq.~\eqref{supp_normalized_frobenius_norm}.  
Also, we define $\Delta t$ as an arbitrary positive constant which is smaller than $1/(2e\tilde{g})$ with $\tilde{g}$ in Eq.~\eqref{supp_tilde_g__def} and satisfies $t/\Delta t \in \mathbb{N}$.
Then, for an arbitrary operators $W_i$ on $i\in \Lambda$ ($\|W_i\|=1$) and its time-evolution of $W_i(t)=e^{iHt} W_i e^{-iHt}$, 
there exists an operator $\tilde{W}^{(t)}_{i[R]}$ which approximates $W_i(t)$ on a region $i[R]$ as follows:
\begin{align}
 \label{supp_ineq:thm_main_poly_light}
\norm{ W_i(t) - \tilde{W}^{(t)}_{i[R]} }_F \le 2^{D-1} C_0 (\Delta t)^{-\alpha+\frac{D+1}{2}} t^{\alpha-\frac{D-1}{2}} R^{-\alpha+D} ,
\end{align} 
where $C_0$ is a constant of $\orderof{1}$ which is defined in Eq.~\eqref{supp_definition_of_C_0} and we assume that $R$ is a multiple of $(t/\Delta t )$.
\end{theorem}
\noindent 
From the above results, we can ensure that the wavefront of the information is restricted in the distance of 
\begin{align}
 R= |t|^{\frac{2\alpha- D+1}{2\alpha-2D}},
\end{align} 
which gives a non-trivial polynomial growth of the OTOC for arbitrary $\alpha>D$.

In the above theorem, we restrict ourselves to an on-site operator $W_i$. 
We can easily extend the theorem to arbitrary operator $W_X$ which are supported on $X\subset \Lambda$.
\begin{corol}[Generalization to arbitrary operators]   \label{supp_corol_main_poly_light}
Let $X$ be an arbitrary subset such that $X \subseteq i[R_0]$. 
Then, for an arbitrary operators $W_X$ ($\|W_X\|=1$) and its time-evolution of $W_X(t)=e^{iHt} W_X e^{-iHt}$, 
there exists an operator $\tilde{W}^{(t)}_{i[R_0+R]}$ which approximates $W_i(t)$ on a region $i[R_0+R]$ as follows:
\begin{align}
 \label{supp_ineq:corol_main_poly_light}
\norm{ W_X(t) - \tilde{W}^{(t)}_{i[R_0+R]} }_F \le 2^{D-1} C_0 (\Delta t)^{-\alpha+\frac{D+1}{2}} t^{\alpha-\frac{D-1}{2}} (R+R_0)^{\frac{D-1}{2}} R^{-\alpha+\frac{D+1}{2}},
\end{align} 
which yields for $D=1$ 
\begin{align}
 \label{supp_ineq:corol_main_poly_light}
\norm{ W_X(t) - \tilde{W}^{(t)}_{i[R_0+R]} }_F \le  \gamma C_0 (\Delta t)^{-\alpha+1} t^{\alpha}  R^{-\alpha+1}.
\end{align} 
\end{corol}

\section{Connection of unitary time evolution}
For readers' convenience, we show the outline of our proof technique again.   
A central technique in our proof is a connection of unitary time evolution addressed in Ref.~\cite{Kuwahara_2016_njp}. 
Following Ref.~\cite{Kuwahara_2016_njp}, we decompose the time to $t/\Delta t$ pieces, and define
\begin{align}
t_m := m \Delta t ,\quad t_{m_t} := t,
\end{align}
where $m_t= t/\Delta t$.
For a fixed $R$ and $i\in \Lambda$, we also define the lengths $\Delta r$, $R_m$ and the subset $X_m$ as follows:
\begin{align}
\label{supp_Choice_Delta_r_X_m}
\Delta r :=R/m_t   ,  \quad X_m:= i[m\Delta r]. 
\end{align}
By using these notations, we approximate $W_i(t_m)$ onto the subset $X_m$. 
For the approximation, we adopt the following recursive procedure.
For $m=1$, we define 
\begin{align}
W_{X_1}^{(1)} := W_i(\Delta t,X_1) ,
\end{align}
where we use the notation of Eq.~\eqref{supp_def:W_X_local_approx}.
Note that $W_i(t,X_1)$ is now supported on the subset $X_1$.
For $m=2$, we define 
\begin{align}
W_{X_2}^{(2)} := W_{X_1}^{(1)}(\Delta t, X_2) . 
\end{align}
We then obtain the approximation error as
\begin{align}
\norm{W_i(2\Delta t) - W_{X_2}^{(2)} }_p& \le  \norm{ W_i(2\Delta t) -W_{X_1}^{(1)}(\Delta t) + W_{X_1}^{(1)}(\Delta t)  - W_{X_2}^{(2)} }_p \notag \\
&\le \norm{ W_i(\Delta t) -W_{X_1}^{(1)}}_p + \norm{W_{X_1}^{(1)}(\Delta t)  - W_{X_2}^{(2)} }_p,
\end{align}
where we use the triangle inequality and the unitary invariance for the Schatten-$p$ norm. 
By repeating this procedure, we obtain 
 \begin{align}
 \label{supp_unitary_connect_upper_bound}
\norm{W_i(m_t\Delta t) - W_{X_{m_t}}^{(m_t)} }_p& \le  \sum_{m=0}^{m_t-1} \norm{ W_{X_m}^{(m)} (\Delta t) - W_{X_{m+1}}^{(m+1)}  }_p,
\end{align}
where we define $W_{X_0}^{(0)}:= W_i$ and $W_{X_j}^{(j)} := W_{X_{j-1}}^{(j-1)}(\Delta t, X_j) $.

As the simplest exercise, let us consider the case of $p=\infty$, which gives the standard operator norm.
By using the Hastings-Koma bound~\eqref{supp_thm_Hastings_ineq} (see~\cite{ref:Hastings2006-ExpDec}), we can obtain 
  \begin{align}
  \label{supp_Hastings_Koma_bound}
\norm{  W_{X_m}^{(m)} (\Delta t) - W_{X_{m+1}}^{(m+1)}  }_\infty \le \frac{2c_1\gamma}{(\alpha-D-1)^2} |\partial X_m| e^{c_2\Delta t} (\Delta r)^{-\alpha+D+1}  .
\end{align}
Thus, by using $|\partial X_m| \le |\partial i[R]|\le  \gamma R^{D-1}$ from \eqref{supp_parameter_gamma_X} and $\Delta r= \Delta t(R/t)$, we obtain 
  \begin{align}
\norm{  W_{X_m}^{(m)} (\Delta t) - W_{X_{m+1}}^{(m+1)}  }_\infty \le \frac{2c_1\gamma^2e^{c_2 \Delta t}(\Delta t)^{-\alpha+D+1}}{(\alpha-D-1)^2} R^{2D-\alpha} t^{\alpha-D-1} .
\end{align}
We thus reduce the upper bound~\eqref{supp_unitary_connect_upper_bound} to
 \begin{align}
 \label{supp_derivation_previous_Lieb_Robinson}
\norm{W_i(t) - W_{i[R]}^{(m_t)} }_\infty& \le  \frac{2c_1\gamma^2 e^{c_2 \Delta t} (\Delta t)^{-\alpha+D}}{(\alpha-D-1)^2}  R^{2D-\alpha} t^{\alpha-D} ,
\end{align}
where we use $m_t=t/\Delta t$. By taking $\Delta t$ as an $\orderof{1}$ constant, we have 
  \begin{align}
\frac{2c_1\gamma^2 e^{c_2 \Delta t} (\Delta t)^{-\alpha+D}}{(\alpha-D-1)^2}  = \orderof{1},
\end{align}
and hence the upper bound~\eqref{supp_derivation_previous_Lieb_Robinson} gives the state-of-the-art ``polynomial light cone'' of $R\approx t^{\frac{\alpha-D}{\alpha-2D}}$, which has been obtained in Ref.~\cite{PhysRevX.9.031006,tran2019locality}.
We note that the result above is more general in the sense that we \textit{do not} need the few-body interactions of the Hamiltonian.
This is because the Hastings-Koma's results only assume the polynomial decay of the interactions~\cite[Assumption~2.1]{ref:Hastings2006-ExpDec}.

\subsection{Local approximation after short time: operator norm}

We here prove the inequality~\eqref{supp_Hastings_Koma_bound}. 
As a useful previous result, we show the theorem by Hastings and Koma~\cite{ref:Hastings2006-ExpDec}: 
\begin{theorem}[Lieb-Robinson bound in long-range interacting systems]  \label{supp_thm_Hastings}
Let $H$ be a Hamiltonian~\eqref{supp_def:Ham} satisfying Assumption~\ref{supp_lem:power_law_k_local}. 
Then, for arbitrary operators $W_X$ and $W_Y$ ($\|W_X\|=\|W_Y\|=1$) defined on subsets $X$ and $Y$, respectively,
 the time-evolved operator $W_X(t):=e^{iHt} W_X e^{-iHt}$ approximately commutes with $W_Y$ as follows:
\begin{align}
\label{supp_thm_Hastings_ineq}
\| [W_X(t), W_Y] \| \le c_1|X| \cdot |Y| \frac{e^{c_2 t}}{(\dist_{X,Y}+1)^\alpha},
\end{align} 
where $c_1$ and $c_2$ are constants of $\orderof{1}$ depending on $J_0$, $\alpha$ and $D$.
\end{theorem}
By using the above theorem and the inequality~\eqref{supp_BHV_bound}, we are going to derive the approximation error of $\| W_X(t)- W_X(t, X[r]) \| $. 
Unfortunately, Theorem~\ref{supp_thm_Hastings} cannot be directly applied to~\eqref{supp_BHV_bound} since $|X[r]^\co|$ is infinitely large in the limit of $n\to \infty$.

For the derivation of the inequality~\eqref{supp_Hastings_Koma_bound}, we obtain the upper bound as follows: 
\begin{lemma}[Local approximation]  \label{supp_thm_local_approximation}
Let $W_X$ be an arbitrary operator on a subset $X$ such that $\|W_X\|=1$.
Then, $W_X(t)$ is approximated by $W_X(t,X[r])$ in Eq.~\eqref{supp_def:W_X_local_approx} as follows:
\begin{align}
\label{supp_improved_BHV_bound}
\| W_X(t)- W_X(t, X[r]) \| \le 2c_1e^{c_2 t} \sum_{i\in X[r]^\co} \sum_{i'\in X} (\dist_{i,i'}+1)^{-\alpha} ,
\end{align}
where $c_1$ and $c_2$ has been defined in Theorem~\ref{supp_thm_Hastings}. 
\end{lemma}

\textit{Proof of Lemma~\ref{supp_thm_local_approximation}.}
We follows the proof in Ref.~\cite{PhysRevX.10.031010} (see Theorem~4 in Supplementary material there). 
We first note that the partial trace in Eq.~\eqref{supp_def:W_X_local_approx} is described by using the random unitary operators:
\begin{align}
W_X(t,\tilde{X})  = \int d\mu(U_{i_1}) \int d\mu(U_{i_2})  \cdots \int d\mu(U_{i_{n_0}})  U_{\tilde{X}^\co}^\dagger W_X(t)  U_{\tilde{X}^\co},
\end{align}
with $\mu(U_i)$ ($i\in \Lambda$) the Haar measure for the unitary operators on $i$, where we define $\tilde{X}^\co=\{i_s\}_{s=1}^{n_0}$ ($n_0:=|\tilde{X}^\co|$) and $U_{\tilde{X}^\co}:=\prod_{i\in \tilde{X}^\co} U_i$.
By applying the above notation to $\tilde{X}=X[r]$, we obtain 
\begin{align}
\| W_X(t) - W_X(t,X[r])\| & \le  \left\| W_X(t)  - \int d\mu(U_{i_1}) \int d\mu(U_{i_2})  \cdots \int d\mu(U_{i_{n_0}})  U_{\tilde{X}^\co}^\dagger W_X(t)  U_{\tilde{X}^\co} \right\|   \notag \\
&\le \sum_{i\in X[r]^\co} \sup_{U_i}\|[W_X(t), U_i ]\| =  \sum_{i\in X[r]^\co} \sup_{U_i}\|[W_X, U_i(-t)]\| . 
\label{supp_error_local_region}
\end{align}

Furthermore, we reduce the commutator norm $\|[W_X, U_i(-t)]\|$ to the following form:
\begin{align}
\label{supp_error_local_region_unitary_0}
\|[W_X, U_i(-t)]\|  = \|[W_X, U_i(-t) - U_i(-t, X^\co) +  U_i(-t, X^\co)  ]\| \le 2 \|W_X\| \cdot \|U_i(-t) - U_i(-t, X^\co) \|,
\end{align}
where we use $[W_X, U_i(-t, X^\co) ]=0$. 
By applying the same inequality as~\eqref{supp_error_local_region} to \eqref{supp_error_local_region_unitary_0}, we have 
\begin{align}
\|U_i(-t) - U_i(-t, X^\co) \|&\le \sum_{i'\in X} \sup_{U_{i'}}\|[U_i(-t) , U_{i'}]\| \le c_1e^{c_2 t} \sum_{i'\in X}  (\dist_{i,i'}+1)^{-\alpha}, 
\label{supp_error_local_region_unitary}
\end{align}
where we use Theorem~\ref{supp_thm_Hastings} in the last inequality. 
By combining the inequalities~\eqref{supp_error_local_region}, \eqref{supp_error_local_region_unitary_0} and \eqref{supp_error_local_region_unitary}, we obtain the main inequality~\eqref{supp_improved_BHV_bound}.
This completes the proof. $\square$

{~}\\

Also, for the summation with respect to $i$ and $i'$, we can derive the following lemma:
\begin{lemma}  \label{supp_thm_sum_i_i'}
Let us consider the case where the subset $X$ is given by $i[r_0]$ for $\forall r_0\in \mathbb{N}$. 
Then, we obtain the upper bound as 
  \begin{align}
\sum_{i\in X} \sum_{i'\in X[r]^\co} (\dist_{i,i'}+1)^{-\alpha} 
&\le \frac{\gamma |\partial X| r^{-\alpha+D+1}}{(\alpha-D-1)^2} ,
\end{align}
\end{lemma}

{~}\\

\textit{Proof of Lemma~\ref{supp_thm_sum_i_i'}.}
By using the notation of $(\partial X)_s$ in Eq.~\eqref{supp_notation_partial_X_s}, we can obtain 
  \begin{align}
\sum_{i\in X} \sum_{i'\in X[r]^\co} (\dist_{i,i'}+1)^{-\alpha} &
\le \sum_{s=0}^\infty \sum_{i\in (\partial X)_{-s}} \sum_{i'\in \Lambda: \dist_{i,i'} > r+s} (\dist_{i,i'}+1)^{-\alpha} .
\end{align}
By using the inequality~\eqref{supp_ineq_sum_decay_func} with $a=\alpha$, the summation with respect to $i'$ is bounded from above by
\begin{align}
\sum_{i'\in \Lambda: \dist_{i,i'} > r+s} (\dist_{i,i'}+1)^{-\alpha} \le \frac{\gamma}{\alpha-D} (r+s+1)^{-\alpha+D},
\end{align}
which yields 
  \begin{align}
\sum_{i\in X} \sum_{i'\in X[r]^\co} (\dist_{i,i'}+1)^{-\alpha} 
&\le \frac{\gamma}{\alpha-D} \sum_{s=0}^\infty \sum_{i\in (\partial X)_{-s}} (r+s+1)^{-\alpha+D}
\le  \frac{\gamma |\partial X| r^{-\alpha+D+1}}{(\alpha-D-1)(\alpha-D)} 
\le \frac{\gamma |\partial X| r^{-\alpha+D+1}}{(\alpha-D-1)^2} ,
\end{align}
where we use $ |(\partial X)_{-s}| \le  |\partial X|$ for $X=i[r_0]$ and 
\begin{align}
\label{supp_ineq_s_sum_integral}
\sum_{s=0}^\infty (r+s+1)^{-\alpha+D} \le\int_r^\infty x^{-\alpha+D} dx 
\le \frac{r^{-\alpha+D+1}}{\alpha-D-1} .
\end{align}
This completes the proof of Lemma~\ref{supp_thm_sum_i_i'}. $\square$

{~}\\

By combining Lemmas~\ref{supp_thm_local_approximation} and \ref{supp_thm_sum_i_i'}, we immediately obtain the inequality~\eqref{supp_Hastings_Koma_bound}.

\section{Proof of our main Theorem~\ref{supp_thm_main_poly_light}}

The proof is based on the inequality~\eqref{supp_unitary_connect_upper_bound} with $p=2$, which gives
 \begin{align}
 \label{supp_unitary_connect_upper_bound_F}
\norm{W_i(m_t\Delta t) - W_{X_{m_t}}^{(m_t)} }_F& \le  \sum_{m=0}^{m_t-1} \norm{ W_{X_m}^{(m)} (\Delta t) - W_{X_{m+1}}^{(m+1)}  }_F.
\end{align}
Here, our key technical ingredient is the following theorem:
\begin{theorem}[Local approximation after short-time evolution]  \label{supp_thm_local_approximation_frobenius}
Let $W_X$ be an arbitrary operator on a subset $X$ with $\|W_X\|=1$.
Then, for $|t|\le 1/(2e\tilde{g})$ (see Eq.~\eqref{supp_tilde_g__def} for the definition of $\tilde{g}$), $W_X(t)$ is approximated by $W_X(t,X[r])$ in Eq.~\eqref{supp_def:W_X_local_approx} as follows:
\begin{align}
\label{supp_Ineq:thm_local_approximation_frobenius}
\| W_X(t)- W_X(t, X[r]) \|_F \le C_0 |t|  \sqrt{\gamma^{-1}  |(\partial X)_{r/2} |  r^{-2\alpha+D+1}},
\end{align}
where we use the notation in Eq.~\eqref{supp_notation_partial_X_s} and $C_0$ is a constant of $\orderof{1}$ which is defined as 
\begin{align}
\label{supp_definition_of_C_0}
C_0 := \frac{2^{\alpha-D/2+2}J_0 \gamma}{2\alpha-D-1} +\frac{80 \tilde{g}\gamma 2^{(3\alpha-D)/2} }{\alpha-D} \sqrt{2\alpha-2D+\gamma}    .
\end{align}
\end{theorem}
\noindent 
We here apply the above theorem with $r=\Delta r$ to $\norm{ W_{X_m}^{(m)} (\Delta t) - W_{X_{m+1}}^{(m+1)}  }_F$.
From $X_m \subseteq i[R]$, we have 
\begin{align}
|(\partial X_m)_{\Delta r/2}| \le  |(\partial i[R])_{R}| =  |(\partial i[2R])| \le \gamma (2R)^{D-1} ,\quad \Delta r= \Delta t(R/t), 
\end{align}
and hence
\begin{align}
\norm{ W_{X_m}^{(m)} (\Delta t) - W_{X_{m+1}}^{(m+1)}  }_F \le 2^{D-1} C_0(\Delta t)^{-\alpha+\frac{D+3}{2}} t^{\alpha-\frac{D+1}{2}} R^{-\alpha+D} .
\end{align}
From $m_t=t/\Delta t$, the above inequality reduces the inequality~\eqref{supp_unitary_connect_upper_bound} to 
\begin{align}
\norm{W_i(m_t\Delta t) - W_{X_{m_t}}^{(m_t)} }_F& \le 2^{D-1} C_0 (\Delta t)^{-\alpha+\frac{D+1}{2}} t^{\alpha-\frac{D-1}{2}} R^{-\alpha+D} . 
\end{align}
We therefore prove Theorem~\ref{supp_thm_main_poly_light}. 
$\square$

The proof of Corollary~\ref{supp_corol_main_poly_light} is almost the same. 
The only difference is that we choose $X_m$ in Eq.~\eqref{supp_Choice_Delta_r_X_m} as  
\begin{align}
X_m:= i[R_0+ m\Delta r]. 
\end{align}
Then, we have 
\begin{align}
|(\partial X_m)_{\Delta r/2}| \le |(\partial i[2R+R_0])| \le 2^{D-1}\gamma (R+R_0)^{D-1},
\end{align}
which makes 
\begin{align}
\norm{ W_{X_m}^{(m)} (\Delta t) - W_{X_{m+1}}^{(m+1)}  }_F \le 2^{D-1} C_0(\Delta t)^{-\alpha+\frac{D+3}{2}} t^{\alpha-\frac{D+1}{2}} (R+R_0)^{\frac{D-1}{2}} R^{-\alpha+\frac{D+1}{2}} , 
\end{align}
which yields the inequality~\eqref{supp_ineq:corol_main_poly_light}. This completes the proof of Corollary~\ref{supp_corol_main_poly_light}. $\square$

\section{Proof of Theorem~\ref{supp_thm_local_approximation_frobenius}: Short-time Lieb-Robinson bound for the Frobenius norm}


Let $Y$ be a subset of $Y=X[r]^\co$. 
For the proof, we utilize the Baker-Campbell-Hausdorff expansion as 
\begin{align}
W_X(t) = \sum_{m=0}^\infty \frac{(it)^m}{m!} \ad_H^m (W_X) , 
\end{align}
where $\ad$ means the commutator, namely $\ad_H(\cdot):= [H,\cdot]$. 
However, the norm of $\ad_H^m (W_X)$ usually scales as $\orderof{|X|^m}$ (see \cite[Lemma~3.1]{KUWAHARA201696}), 
and hence for $|X|\gg 1$, we cannot obtain meaningful upper bound.

In order to overcome this problem, we utilize the standard differential recursion approach.
We start from the inequality~\eqref{supp_BHV_bound}: 
\begin{align}
\label{supp_BHV_bound_start}
\| W_X(t)- W_X(t, X[r]) \|_F \le \int d\mu(U_Y) \| W_X(t)- U_Y^\dagger W_X(t) U_Y \|_F \le \sup_{U_Y} \| [ W_X(t), U_Y] \|_F .
\end{align}
For the commutator norm $\| [ W_X(t), U_Y] \|_F$, we can derive the following inequality:
\begin{align}
\label{supp_LR_bound_start}
\frac{d}{dt} \| [ W_X(t), U_Y] \|_F \le 2\|W_X\| \cdot \| [H_X (t), U_Y]\|_F  ,
\end{align}
where we define $H_X$ as 
\begin{align}
H_X: = \sum_{Z:Z\cap X\neq \emptyset} h_Z .
\end{align}
The proof of this inequality is followed by the same approach  as in~\cite[Inequality~(A.10)]{ref:Hastings2006-ExpDec}. 
Therefore, we need to consider the quasi-locality of the time evolution of $k$-local Hamiltonian [i.e., $H_X(t)$] instead of the original time evolution $W_X(t)$.

For the estimation of $\| [H_X (t), U_Y]\|_F$, we first decompose the Hamiltonian $H_X$ as follows: 
\begin{align}
H_X = H_X^{(\le r/2)} + H_X^{(> r/2)},
\end{align}
where $H_X^{(\le r/2)}$ and $H_X^{(> r/2)}$ include the interactions of $h_Z$ such that $Z\subseteq X[r/2]$ and $Z\cap X[r/2]^\co\neq \emptyset$, respectively.
We first consider $\| [H_X^{(> r/2)} (t), U_Y]\|_F $. By using the H\"older inequality~\eqref{supp_sup_generalized_Holder_ineq_operator_norm}, we have $\|O_1O_2\|_F \le \|O_1\| \cdot \|O_2\|_F$ and hence 
\begin{align}
\label{supp_first_term_r/2>}
\norm{ [H_X^{(> r/2)} (t), U_Y]}_F \le   2 \norm{H_X^{(> r/2)}}_F   = 2\norm{ \sum_{Z : Z\cap X[r/2]^\co\neq \emptyset} h_Z  }_F =
2\sqrt{\sum_{Z\cap X[r/2]^\co\neq \emptyset } \ftr(h_Z^2)} ,
\end{align}
where we use the unitary invariance of the Frobenius norm, namely $\| H_X^{(> r/2)} (t)\|_F=\| H_X^{(> r/2)}\|_F$.
We remind that $\ftr(\cdots)$ has been defined in Eq.~\eqref{supp_normalized_frobenius_norm}.
By using the following notation of 
\begin{align}
H_{i,i'} = \sum_{Z:Z\supset \{i,i'\}} h_Z ,
\end{align}
we have 
\begin{align}
 \label{supp_first_term_r/2>_upp_0}
\norm{ [H_X^{(> r/2)} (t), U_Y]}_F^2
\le 4\sum_{s=0}^\infty \sum_{i\in (\partial X)_{-s}} \sum_{i': \dist_{i,i'}>s+r/2} \sum_{Z :Z\supset \{i,i'\}} \ftr(h_{Z}^2) 
= 4\sum_{s=0}^\infty \sum_{i\in (\partial X)_{-s}} \sum_{i': \dist_{i,i'}>s+r/2}  \ftr(H_{i,i'}^2).
\end{align}
The condition~\eqref{supp_alternative_basic_assump_power} for the Hamiltonian gives $\ftr(H_{i,i'}^2) \le \|H_{i,i'}\|^2 \le J_0^2 (\dist_{i,i'}+1)^{-2\alpha}$, 
and hence the inequality~\eqref{supp_ineq_sum_decay_func} reduces the above inequality to 
 \begin{align}
 \label{supp_first_term_r/2>_upp}
4 J_0^2\sum_{s=0}^\infty \sum_{i\in (\partial X)_{-s}} \sum_{i': \dist_{i,i'}>s+r/2} (\dist_{i,i'}+1)^{-2\alpha}  
&\le  \frac{4J_0^2\gamma}{2\alpha-D}|\partial X| \sum_{s=0}^\infty (s+r/2+1)^{-2\alpha+D} \notag \\
&\le   \frac{2^{2\alpha-D+2}J_0^2\gamma}{(2\alpha-D-1)^2} |\partial X|  r^{-2\alpha+D+1},
\end{align}
where we use a similar inequality to~\eqref{supp_ineq_s_sum_integral} in the last inequality. 
By applying the inequality~\eqref{supp_first_term_r/2>_upp} to \eqref{supp_first_term_r/2>_upp_0}, we obtain 
\begin{align}
\label{supp_first_term_r/2>_final_form}
\norm{ [H_X^{(> r/2)} (t), U_Y]}_F \le  C_1\sqrt{ \gamma^{-1}|\partial X|  r^{-2\alpha+D+1}},
\end{align}
where $C_1:=  2^{\alpha-D/2+1}J_0 \gamma / (2\alpha-D-1)$ which is a constant of $\orderof{1}$. 

On the other hand, the estimation of $\norm{ [H_X^{(\le r/2)} (t), U_Y]}_F$ is much more intricate.
Most of the following discussions are devoted to prove the following proposition:
\begin{prop} \label{supp_prop:time_evolve_short_time}
As long as $t$ is smaller than $1/(2e\tilde{g})$ ($|t|\le 1/(2e\tilde{g})$), we obtain the upper bound of $\norm{ [H_X^{(\le r/2)} (t), U_Y]}_F$ as 
\begin{align}
\label{supp_first_term_<=r/2_final_form}
\norm{ [H_X^{(\le r/2)} (t), U_Y]}_F \le C_2 \sqrt{\gamma^{-1}  |(\partial X)_{r/2} |   r^{-2\alpha+D+1}},
\end{align}
where $C_2$ is given by
\begin{align}
C_2 =\frac{40 \tilde{g}\gamma 2^{(3\alpha-D)/2} }{\alpha-D} \sqrt{2\alpha-2D+\gamma} .
\end{align}
\end{prop}
\noindent
By applying the inequalities~\eqref{supp_first_term_r/2>_final_form} and \eqref{supp_first_term_<=r/2_final_form} to \eqref{supp_LR_bound_start}, we obtain
\begin{align}
\frac{d}{dt} \| [ W_X(t), U_Y] \|_F \le 2\|W_X\|  (C_1+C_2) \sqrt{\gamma^{-1}  |(\partial X)_{r/2} |  r^{-2\alpha+D+1}},
\end{align}
where we use $  |\partial X | \le  |(\partial X)_{r/2} | $.
By taking the integral with respect to $t$, we thus prove the main inequality~\eqref{supp_Ineq:thm_local_approximation_frobenius} from the inequality~\eqref{supp_BHV_bound_start}.
This completes the proof of Theorem~\ref{supp_thm_local_approximation_frobenius}. $\square$


\section{Proof of Proposition~\ref{supp_prop:time_evolve_short_time}}

\subsection{Expansion of the Hamiltonian by the Pauli bases}

For the proof, we first introduce the notation of the expansion of the Hamiltonian with respect to the Pauli bases. 
We denote local terms $\{h_Z\}_{Z\subset \Lambda}$ in the Hamiltonian as follows:
\begin{align}
 h_Z=\sum_{q=1}^{3^{|Z|}} J_{Z,q} \Pau_{Z,q},
\end{align}
where $\Pau_{Z,q}$ is given by a product of Pauli's matrices on the subset $Z$, such as $\Pau_{Z,q}= \sigma_{i_1}^x \sigma_{i_2}^x \sigma_{i_3}^y \sigma_{i_4}^z$ with $Z=\{i_1,i_2,i_3,i_4\}$. We denote $\Pau_{Z,0}=0$ for arbitrary $Z\subset \Lambda$.
Then, the original condition implies 
\begin{align}
\sum_{Z:Z\supset \{i,i'\}} \sqrt{ \sum_{q=1}^{3^{|Z|}} | J_{Z,q} |^2} \le  \sum_{Z:Z\supset \{i,i'\}} \|h_Z\| \le  J_0  (\dist_{i,i'}+1)^{-\alpha},
\end{align}
and hence 
\begin{align}
\sum_{Z:Z\supset \{i,i'\}} \sum_{q=1}^{3^{|Z|}} | J_{Z,q} |  \le  3^{k/2} J_0  (\dist_{i,i'}+1)^{-\alpha} =: J (\dist_{i,i'}+1)^{-\alpha} \quad {\rm with} \quad J:=3^{k/2} J_0, 
\end{align}
where we use $\|h_Z\| \ge \|h_Z\|_F$, $\sqrt{\sum_{q=1}^{3^{|Z|}} | J_{Z,q} |^2} \ge 3^{-|Z|/2} \sum_{q=1}^{3^{|Z|}} | J_{Z,q} |$\footnote{It can be easily obtained from the convexity of $f(z)=z^2$. For arbitrary $\{x_j\}_{k=1}^m$ ($x_j>0$), we have 
\begin{align}
f( [x_1+x_2 +\cdots +x_n]/n) \le \frac{f(x_1)+f(x_2) +\cdots +f(x_n)}{n}.
\end{align}
By taking the square root of the above inequality and letting $\{x_j\}_{k=1}^m \to \{|J_{Z,q}|\}_{q=1}^{3^{|Z|}}$, we can derive the desired upper bound.
} and $|Z|\le k$.

On the product of Pauli's operator, we can obtain the following convenient relations:
\begin{align}
[ \Pau_{Z,q} , \Pau_{Z',q'} ] = 2\eta_{(Z,q),(Z',q')} \Pau_{\tilde{Z},\tilde{q}} \label{supp_commutator_pau_Z_q}
\end{align}
with $\tilde{Z} \subseteq Z\cup Z'$, where the quantity $\eta_{(Z,q),(Z',q')}$ has a value $-1$ or $1$ and we choose $\tilde{q}=0$ when $[ \Pau_{Z,q} , \Pau_{Z',q'} ] =0$, namely $\Pau_{Z,0}=0$ for $\forall Z \subseteq \Lambda$. 
For example, for $\Pau_{Z,q}=\sigma_1^x\sigma_2^x\sigma_3^y$ and $\Pau_{Z',q'}=\sigma_3^z\sigma_4^z$, we have 
$
[\Pau_{Z,q},\Pau_{Z',q'}]=2\sigma_1^x\sigma_2^x\sigma_3^x\sigma_4^z
$
and $\tilde{Z}=\{1,2,3,4\}=Z\cup Z'$.
We notice that $\tilde{Z}$ is not usually equal to $Z\cup Z'$; for example, for $\Pau_{Z,q}=\sigma_1^x\sigma_2^x\sigma_3^x\sigma_4^x$ and 
$\Pau_{Z',q'}=\sigma_1^x\sigma_2^x\sigma_3^x\sigma_4^y$, we have $[\Pau_{Z,q},\Pau_{Z',q'}] =2 \sigma_4^z$ and $\tilde{Z}=\{4\} \subset Z\cup Z'$.

 \begin{figure}[tt]
\centering
\includegraphics[clip, scale=0.4]{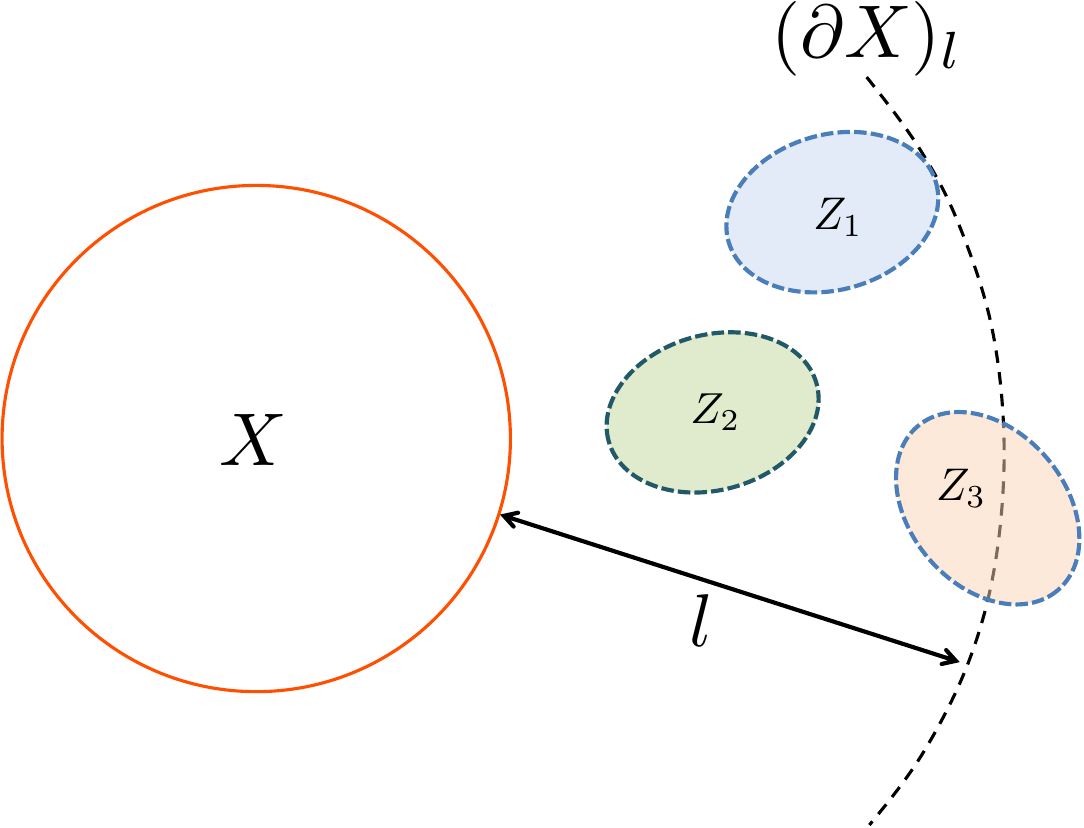}
\caption{Schematic picture of the set $\Set_l$ defined as in Eq.~\eqref{supp_Set_S_r}. Here, $Z_1$ is included in $\Set_l$ as it has an overlap with $(\partial X)_l$ and $Z_1 \subseteq \bigcup_{s\le l} (\partial X)_s$. Then, $Z_2$ is not included in $\Set_l$ because $Z_2 \cap (\partial X)_l = \emptyset$. The subset $Z_3$ is also not included in $\Set_l$ because $Z_3$ is not included in $\bigcup_{s\le l} (\partial X)_s$, namely $Z_3 \cap  \bigcup_{s>l} (\partial X)_s \neq \emptyset$.}
\label{supp_fig_partial_X_s}
\end{figure}

For the convenience, we also define the following ``one-site energy'' as $g$:
\begin{align}
\max_{i\in \Lambda} \sum_{\substack{(Z,q): Z\ni i}} |J_{Z,q} |  \le g \label{supp_g_extensive},
\end{align}
where $g$ is a constant of $\orderof{1}$. 
We notice that an explicit form of $g$ is derived by using the same inequality as \eqref{supp_ineq_sum_decay_func}: 
\begin{align}
\label{supp_g_extensive_explicit}
\sum_{\substack{(Z,q): Z \ni i}} |J_{Z,q} |  &\le 
\sum_{i'\in \Lambda} \sum_{\substack{(Z,q): Z \supset \{i,i'\}}} |J_{Z,q} |  \le \sum_{i'\in \Lambda} \frac{J}{(\dist_{i,i'}+1)^{\alpha}}   \notag \\
&\le \sum_{s=0}^\infty \sum_{i'\in \Lambda: \dist_{i,i'}=s} J (s+1)^{-\alpha}  \le \frac{\alpha-D+1}{\alpha-D} J \gamma ,
\end{align}
which is finite as long as $\alpha>D$.

Finally, in order to denote $H_X^{(\le r/2)}$, we define $\Set_{l}$ as a set of $\{Z\}_{|Z|\le k}$ satisfying 
\begin{align}
\label{supp_Set_S_r}
&\Set_{l} := \left \{Z\subset \Lambda\  \biggl | \ |Z| \le k, \ Z\cap (\partial X)_l \neq \emptyset, \ Z \subseteq \bigcup_{s\le l} (\partial X)_s \right \}, \quad \Set_{\le l}:=\bigsqcup_{s\le l} \Set_s.  
\end{align}
Then, all the interactions in $H_X^{(\le r/2)}$ is described by $\{h_Z\}_{Z\in \Set_{\le r/2}}$.
Hence, by using $\Set_{\le r/2}$, we write $H_X^{(\le r/2)}$ as 
\begin{align}
\label{supp_notation_H_X_le_r/2}
H_X^{(\le r/2)} =\sum_{l=-\infty}^{r/2} \sum_{Z\in \Set_{l} } h_Z = \sum_{Z\in \Set_{\le r/2}}  \sum_{q=1}^{3^{|Z|}} J_{Z,q} \Pau_{Z,q}.
\end{align}

\subsection{Proof outline}

We here aim estimate the Frobenius norm of 
\begin{align}
 \norm{ H_X^{(\le r/2)}(t)- H_X^{(\le r/2)}(t,X[r]) }_F,
\end{align}
which gives an upper bound of $\norm{ [H_X^{(\le r/2)} (t), U_Y]}_F$ ($Y=X[r]^\co$) as follows:
\begin{align}
\label{supp_upp_local_approx_commute}
\norm{ [H_X^{(\le r/2)} (t), U_Y]}_F &\le  \norm{ [ H^{(\le r/2)}_X(t)- H^{(\le r/2)}_X(t,X[r]), U_Y]}_F  + \norm{ [H^{(\le r/2)}_X(t,X[r]), U_Y]}_F   \notag \\
&\le 2 \|U_Y\| \cdot \norm{ H^{(\le r/2)}_X(t)- H^{(\le r/2)}_X(t,X[r])}_F = 2 \norm{ H^{(\le r/2)}_X(t)- H^{(\le r/2)}_X(t,X[r])}_F,
\end{align}
where we use $[H_X^{(\le r/2)}(t,X[r]), U_Y]=0$ and $\|O_1O_2\|_F \le \|O_1\| \cdot \|O_2\|_F$ from the inequality~\eqref{supp_sup_generalized_Holder_ineq_operator_norm}.
In the following, we calculate an upper bound of $\norm{ H_X^{(\le r/2)}(t)- H_X^{(\le r/2)}(t,X[r])}_F$.

For the purpose, we consider a time evolution of $J_{Z_0,q_0}\Pau_{Z_0,q_0}$ as 
\begin{align}
J_{Z_0,q_0} \Pau_{Z_0,q_0} (t) = \sum_{m=0}^\infty \frac{(it)^m}{m!} J_{Z_0,q_0}\ad_H^m (\Pau_{Z_0,q_0}) . \label{supp_Pau_Z_q_t_start}
\end{align}
Here, $J_{Z_0,q_0}\ad_H^m (\Pau_{Z_0,q_0})$ is composed of the following multi-commutators: 
\begin{align}
\br{J_{Z_0,q_0} J_{Z_1,q_1} J_{Z_2,q_2} \cdots J_{Z_m,q_m}}  \ad_{\Pau_{Z_m,q_m}}  \cdots \ad_{\Pau_{Z_2,q_2}}  \ad_{\Pau_{Z_1,q_1}} (\Pau_{Z_0,q_0}) .
\end{align}
We here define 
\begin{align}
w:=((Z_0,q_0),(Z_1,q_1),(Z_2,q_2),\ldots,(Z_m,q_m))
\end{align}
as a string of $\{(Z,q)\}_{|Z|\le k}$.  
Then, from Eq.~\eqref{supp_commutator_pau_Z_q}, the multi-commutator $\ad_{\Pau_{Z_m,q_m}}  \cdots \ad_{\Pau_{Z_2,q_2}}  \ad_{\Pau_{Z_1,q_1}} (\Pau_{Z_0,q_0})$ reduces to the following form:
\begin{align}
 \ad_{\Pau_{Z_m,q_m}}  \cdots \ad_{\Pau_{Z_2,q_2}}  \ad_{\Pau_{Z_1,q_1}} (\Pau_{Z_0,q_0}) = 
 2^m \eta_{w} \Pau_{\Lambda_w,q_w}, 
\label{supp_Def_Pau_V_Z_q_w}
\end{align}
where $\eta_{w}$ has a value $-1$ or $1$, and $q_w$ has a quantity from $0$ to $3^{|\Lambda_w|}$ (remember that $\Pau_{\Lambda_w,0}=0$). 
Note that the subset $\Lambda_w$ and the index $q_w$ are defined by the equation~\eqref{supp_Def_Pau_V_Z_q_w} and non-trivially depend on the choice of $w$. 
Here, the subset $\Lambda_w$ is included in the subset $\overline{\Lambda}_w$  as 
\begin{align}
\label{supp_def:tilde_V___w}
\overline{\Lambda}_w := Z_0 \cup Z_1\cup Z_2 \cup \cdots \cup Z_{m} .
\end{align}

Hence, by using the notation of 
\begin{align}
&J_w :=J_{Z_0,q_0}J_{Z_1,q_1} J_{Z_2,q_2} \cdots J_{Z_m,q_m},
\end{align}
we formally write Eq.~\eqref{supp_Pau_Z_q_t_start} as 
\begin{align}
J_{Z_0,q_0} \Pau_{Z_0,q_0} (t) = \sum_{m=0}^\infty \frac{(2it)^m}{m!} \sum_{\substack{w\in \Omega_{m} \\ (Z_0,q_0):{\rm fixed} }}  J_w  \eta_{w}  \Pau_{\Lambda_w,q_w}  , \label{supp_Pau_Z_q_t_cluster_notation}
\end{align}
where we define the total set of the string $w$ with $|w|=m+1$ as $\Omega_{m}$.

By using the expressions~\eqref{supp_notation_H_X_le_r/2} and \eqref{supp_Pau_Z_q_t_cluster_notation}, we obtain 
\begin{align}
H_X^{(\le r/2)} (t) = \sum_{(Z_0,q_0):Z_0\in \Set_{\le r/2}} J_{Z_0,q_0} \Pau_{Z_0,q_0} (t) =\sum_{m=0}^\infty \frac{(2it)^m}{m!} \sum_{w\in \Omega_{m}: Z_0\in \Set_{\le r/2}}  J_w  \eta_w \Pau_{\Lambda_w,q_w}  , \label{supp_Pau_Z_q_t_cluster_notation_1}
\end{align}
where $\sum_{w\in \Omega_{m}: Z_0\in \Set_{\le r/2}}$ means the summation which picks up all the strings which satisfy $Z_0 \in \Set_{\le r/2}$.
We then obtain 
\begin{align}
H_X^{(\le r/2)}(t)- H^{(\le r/2)}_X(t,X[r])  = \sum_{m=0}^\infty \frac{(2it)^m}{m!} \sum_{\substack{w\in \Omega_{m}: Z_0\in \Set_{\le r/2} \\ \Lambda_w \cap X[r]^\co \neq \emptyset }} J_w   \eta_w\Pau_{\Lambda_w,q_w}   .\label{supp_Pau_Z_q_t_cluster_notation__2}
\end{align}
For $\Pau_{\Lambda_w,q_w}$, we obtain the equation of
\begin{align}
\tr \br{ \Pau_{\Lambda_w,q_w}\Pau_{\Lambda_{w'},q_{w'}}   } = 0  \for \Lambda_w \neq \Lambda_{w'} .
\end{align}
We note that we may have $\Pau_{\Lambda_w,q_w}=\Pau_{\Lambda_{w'},q_{w'}}$ for $w \neq w'$, and hence the condition $w \neq w'$ does not necessarily imply $\tr (\Pau_{\Lambda_w,q_w}\Pau_{\Lambda_{w'},q_{w'}})=0$.
We thus obtain 
\begin{align}
&\norm {H_X^{(\le r/2)}(t)- H^{(\le r/2)}_X(t,X[r])}_F^2   \notag \\
& \le  \sum_{m,m'=0}^\infty \frac{(2|t|)^{m+m'}}{m!m'!}
\sum_{\substack{w\in \Omega_{m}: Z_0\in \Set_{\le r/2} \\ \Lambda_w \cap X[r]^\co \neq \emptyset }}  
\sum_{\substack{w'\in \Omega_{m'}: Z'_0\in \Set_{\le r/2}  \\ \Lambda_{w'} =  \Lambda_w }}   | J_w | \cdot |J_{w'}| \cdot \norm{ \Pau_{\Lambda_w,q_w} \Pau_{\Lambda_{w'},q_{w'}} } , \label{supp_Pau_Z_q_t_cluster_notation_2}
\end{align}
where we use the inequality  of 
\begin{align}
\tilde{\tr} \br{ \Pau_{\Lambda_w,q_w}\Pau_{\Lambda_{w'},q_{w'}}   } \le \norm{\Pau_{\Lambda_w,q_w}\Pau_{\Lambda_{w'},q_{w'}} }. 
\end{align}
Note that the norm $\|\Pau_{\Lambda_w,q_w}\| $ has the binary values of $0$ or $1$.

The remaining task is to estimate the upper bound~\eqref{supp_Pau_Z_q_t_cluster_notation_2}.
We first consider the condition that $\Pau_{\Lambda_w,q_w} \neq 0$. 
From the expression~\eqref{supp_Def_Pau_V_Z_q_w}, we find that each of $Z_j$ ($j\le m$) should satisfy 
\begin{align}
\label{supp_condition_subset_Z_j}
Z_j \cap (Z_0\cup Z_1\cup Z_2 \cup \cdots \cup Z_{j-1}) \neq \emptyset.
\end{align}
Otherwise, $\Pau_{Z_j,q_j}$ and $\ad_{\Pau_{Z_{j-1},q_{j-1}}}  \cdots \ad_{\Pau_{Z_2,q_2}}  \ad_{\Pau_{Z_1,q_1}} (\Pau_{Z_0,q_0})$ commute with each other.
We define $\Omega^\ast_m$ as the set of $w$ which satisfies the condition~\eqref{supp_condition_subset_Z_j} for all $\{Z_j\}_{j=1}^m$, namely  
\begin{align}
\Omega^\ast_m=\left\{((Z_j,q_j))_{j=1}^m \biggl | Z_j \cap (Z_0\cup Z_1\cup Z_2 \cup \cdots \cup Z_{j-1}) \neq \emptyset \for \forall j \in [m] \right\}.
\end{align}
By using the notation of $\Omega^\ast_m$, we obtain
\begin{align}
&\sum_{\substack{w\in \Omega_{m}: Z_0\in \Set_{\le r/2} \\ \Lambda_w \cap X[r]^\co \neq \emptyset }}  
\sum_{\substack{w'\in \Omega_{m'}: Z'_0\in \Set_{\le r/2}  \\ \Lambda_{w'} =  \Lambda_w }}   | J_w | \cdot |J_{w'}| \cdot \norm{ \Pau_{\Lambda_w,q_w} \Pau_{\Lambda_{w'},q_{w'}} }  
\le \sum_{\substack{ w\in \Omega^\ast_m: Z_0\in \Set_{\le r/2}  \\  \Lambda_w \cap X[r]^\co \neq \emptyset }}  \sum_{\substack{ w'\in \Omega^\ast_{m'} \\  \Lambda_{w'}=\Lambda_w }} | J_w | \cdot |J_{w'}| ,
\end{align}
where we use $\| \Pau_{\Lambda_w,q_w}\| \le 1$.
We notice again that in the above summation for $w'$, more than one string $w'$ may satisfy $\Lambda_{w'}=\Lambda_w$. 
We thus reduce the inequality~\eqref{supp_Pau_Z_q_t_cluster_notation_2} to
\begin{align}
\label{supp_Ineq:Frobenius_norm_simplified_form}
\norm {H_X^{(\le r/2)}(t)- H^{(\le r/2)}_X(t,X[r])}_F^2 
&\le  \sum_{m=0}^\infty \frac{(2|t|)^m}{m!}
\sum_{\substack{ w\in \Omega^\ast_m: Z_0\in \Set_{\le r/2}  \\  \Lambda_w \cap X[r]^\co \neq \emptyset }} | J_w |  \sum_{m'=0}^\infty \frac{(2|t|)^{m'}}{m'!}  \sum_{\substack{ w'\in \Omega^\ast_{m'} \\  \Lambda_{w'}=\Lambda_w }}  |J_{w'}|  \notag \\
&\le  \sum_{i\in X[r]^\co}
\sum_{m=0}^\infty \frac{(2|t|)^m}{m!}
\sum_{\substack{ w\in \Omega^\ast_m: Z_0\in \Set_{\le r/2}  \\  \Lambda_w \ni i}} | J_w |  \sum_{m'=0}^\infty \frac{(2|t|)^{m'}}{m'!}  \sum_{\substack{ w'\in \Omega^\ast_{m'} \\  \Lambda_{w'}=\Lambda_w }}  |J_{w'}| .
\end{align}

In the following, we separately treat the summations with respect to $w'$ and $w$, respectively. 
We take the two steps as 
\begin{quote}
 \begin{itemize}
  \item{} In the first step, for a fixed $\Lambda_w$ such that $w \in \Omega^\ast_m$, we take the summation with respect to $w'\in \Omega^\ast_{m'}$ such that $\Lambda_{w'}=\Lambda_w$.  We aim to obtain the inequality of
  \begin{align}
  \label{supp_first_step_inequality}
 \sum_{m'=0}^\infty \frac{(2|t|)^{m'}}{m'!} \sum_{\substack{ w'\in \Omega^\ast_{m'} \\  \Lambda_{w'}=\Lambda_w }}  |J_{w'}|  \le \frac{6 \tilde{g}}{[\diam(\Lambda_w)+1]^\alpha} ,
\end{align}
where $\tilde{g}:=\max(gk, \lambda J)] $ as in Eq.~\eqref{supp_tilde_g__def}. 
 \item{} In the second step, we take the summation with respect to $w\in \Omega^\ast_{m}$ such that $Z_0\in \Set_{\le r/2}$ and $\Lambda_w \ni i$ ($i\in X[r]^\co$), which gives the following upper bound:
 \begin{align}
 \label{supp_aim_eq:estimate the summation with respect to w}
\sum_{\substack{ w\in \Omega^\ast_m :Z_0\in \Set_{\le r/2}\\  \Lambda_w\ni i}} \frac{| J_w |}{[\diam(\Lambda_w)+1]^\alpha} 
&\le  c_2 m!  (m+1)^4\tilde{g}^{m+1} \sum_{l=-\infty}^{r/2} \sum_{i_1 \in (\partial X)_l}  \frac{1}{(\dist_{i,i_1}+1)^{2\alpha}},
 \end{align}
 where $c_2$ is defined as
  \begin{align}
 \label{supp_definition_of_c_2}
c_2 :=2^{\alpha}\br{2 + \frac{\gamma }{\alpha-D} }.
 \end{align}
   \end{itemize}
\end{quote}

By applying the upper bounds~\eqref{supp_first_step_inequality} and \eqref{supp_aim_eq:estimate the summation with respect to w} to 
the inequality~\eqref{supp_Ineq:Frobenius_norm_simplified_form}, we finally obtain 
\begin{align}
\norm {H_X^{(\le r/2)}(t)- H^{(\le r/2)}_X(t,X[r])}_F^2 
&\le 6c_2 \tilde{g}^2 
\sum_{m=0}^\infty (2\tilde{g} |t|)^m (m+1)^4  \sum_{i\in X[r]^\co} \sum_{l=-\infty}^{r/2} \sum_{i_1 \in (\partial X)_l}  \frac{1}{(\dist_{i,i_1}+1)^{2\alpha}} \notag \\
&\le 392 c_2 \tilde{g}^2  \sum_{i\in X[r]^\co}\sum_{l=-\infty}^{r/2} \sum_{i_1 \in (\partial X)_l}  \frac{1}{(\dist_{i,i_1}+1)^{2\alpha}} ,
\end{align}
where the second inequality is derived from $|t|\le 1/(2e\tilde{g})$ and
\begin{align}
\sum_{m=0}^\infty (2\tilde{g} |t|)^m (m+1)^4 \le \sum_{m=0}^\infty e^{-m} (m+1)^4 \approx 65.2478 .
\end{align}
The summations with respect to $i\in X[r]^\co$ and $i_1\in  (\partial X)_l$ are estimated by a similar approach to Lemma~\ref{supp_thm_sum_i_i'}.
We first note that $\dist_{i,i_1} > r-l$ because of $i\in X[r]^\co$ and $i_1 \in (\partial X)_l$. 
Hence, we obtain
\begin{align}
 \sum_{i\in X[r]^\co}\frac{1}{(\dist_{i,i_1}+1)^{2\alpha}} \le  
 \sum_{i\in \Lambda : \dist_{i,i_1}> r-l } \frac{1}{(\dist_{i,i_1}+1)^{2\alpha}}  \le \frac{\gamma}{2\alpha-D} (r-l+1)^{-2\alpha+D} \le \gamma (r-l+1)^{-2\alpha+D} ,
\end{align}
where we use the inequality~\eqref{supp_ineq_sum_decay_func} with $a=2\alpha$ from the second inequality to the third inequality.
Then, by using the inequality of
\begin{align}
\sum_{l=-\infty}^{r/2} (r-l+1)^{-2\alpha+D} \le  \int_{-\infty}^{r/2}(r-x)^{-2\alpha+D} dx \le \frac{(r/2)^{-2\alpha+D+1}}{2\alpha-D-1} \le 
\frac{2^{2\alpha-D}}{\alpha-D}r^{-2\alpha+D+1} ,
\end{align}
we obtain
\begin{align}
 \sum_{i\in X[r]^\co}\sum_{l=-\infty}^{r/2} \sum_{i_1 \in (\partial X)_l}  \frac{1}{(\dist_{i,i_1}+1)^{2\alpha}}&
\le  |(\partial X)_{r/2} |\frac{2^{2\alpha-D}\gamma}{\alpha-D}r^{-2\alpha+D+1}  , 
\end{align}
where we use $|(\partial X)_{l} |\le |(\partial X)_{r/2} |$ for $l\le r/2$.

Thus, we finally obtain 
\begin{align}
\norm {H_X^{(\le r/2)}(t)- H^{(\le r/2)}_X(t,X[r])}_F^2 
&\le   |(\partial X)_{r/2} | \frac{392 c_2 \tilde{g}^2 2^{2\alpha-D}\gamma }{\alpha-D} r^{-2\alpha+D+1} ,
\end{align}
which yields the main inequality~\eqref{supp_first_term_<=r/2_final_form} by using \eqref{supp_upp_local_approx_commute}, where we use the inequality of
\begin{align}
\sqrt{ \frac{392 c_2 \tilde{g}^2 2^{2\alpha-D}\gamma^2 }{\alpha-D} } \le \frac{20 \tilde{g}\gamma 2^{(3\alpha-D)/2} }{\alpha-D} \sqrt{2\alpha-2D+\gamma}.
\end{align}
This completes the proof of Proposition~\ref{supp_prop:time_evolve_short_time}. $\square$

\subsection{Estimation of the summation with respect to $w'$: Proof of \eqref{supp_first_step_inequality}}

\begin{figure}[bb]
\centering
\subfigure[One example for $m=6$]
{\includegraphics[clip, scale=0.7]{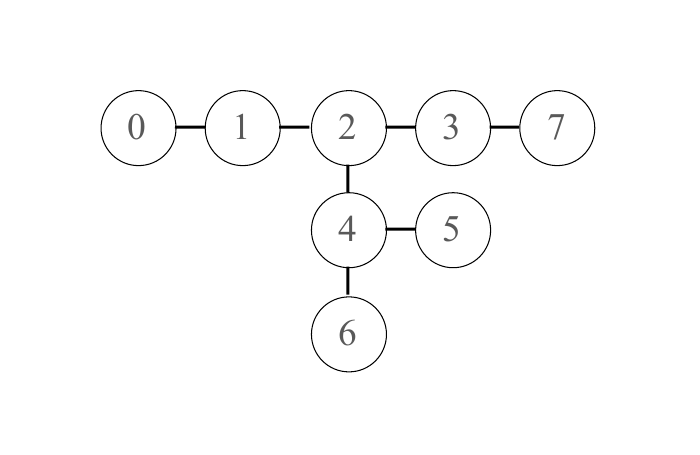}}
\subfigure[All the graphs for $m=3$]
{\includegraphics[clip, scale=0.5]{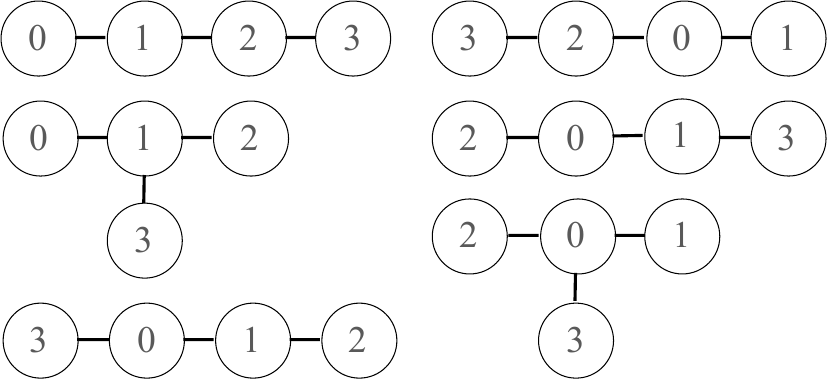}}
\caption{
}
\label{supp_connection_graph}
\end{figure}

For the proof of \eqref{supp_first_step_inequality}, we first consider a fixed $m'$. 
Let us set $\diam( \Lambda_w)=\ell$. 
Then, there exists a pair of sites $\{i,i'\} \subset \Lambda_w$ which satisfies $\dist_{i,i'}=\ell$. 
We thus obtain the following upper bound for the summation: 
  \begin{align}
  \label{supp_2point_include_sum}
\sum_{\substack{ w'\in \Omega^\ast_{m'} \\  \Lambda_{w'}=\Lambda_w }} |J_{w'}|  \le \sum_{\substack{ w'\in \Omega^\ast_{m'} \\  \Lambda_{w'} \ni \{i,i'\}}} |J_{w'}| 
 \le \sum_{\substack{ w'\in \Omega^\ast_{m'} \\  \overline{\Lambda}_{w'} \ni \{i,i'\}}} |J_{w'}|,
\end{align}
where $\overline{\Lambda}_{w'}$ has been defined in Eq.~\eqref{supp_def:tilde_V___w}.
For the above summation, we prove the following lemma:
\begin{lemma} \label{supp_lem:2point_include_sum}
The summation with respect to strings $w\in \Omega^\ast_m$ under the constraint $\overline{\Lambda}_{w} \ni \{i,i'\}$ is upper-bounded by
  \begin{align}
  \label{supp_2point_include_sum_upp}
\sum_{\substack{ w\in \Omega^\ast_{m} \\  \overline{\Lambda}_{w} \ni \{i,i'\}}} |J_{w}|  \le m! (m+1)^2\tilde{g}^{m+1} (\dist_{i,i'}+1)^{-\alpha},
\end{align}
where we use the definition $\tilde{g}:=\max(gk, \lambda J)] $. 
\end{lemma}

\noindent
By applying Lemma~\ref{supp_lem:2point_include_sum} to \eqref{supp_2point_include_sum} with $\dist_{i,i'}=\ell$, we obtain the following upper bound:
 \begin{align}
 \sum_{m'=0}^\infty \frac{(2|t|)^{m'}}{m'!} \sum_{\substack{ w'\in \Omega^\ast_{m'} \\  \Lambda_{w'}=\Lambda_w }} |J_{w'}|   
 &\le \sum_{m'=0}^\infty \frac{(2|t|)^{m'}}{m'!} \sum_{\substack{ w'\in \Omega^\ast_{m'} \\  \overline{\Lambda}_{w'} \ni \{i,i'\}}} |J_{w'}|  \le \tilde{g} (\ell+1)^{-\alpha}  \sum_{m'=0}^\infty  (m'+1)^2 (2\tilde{g} |t|)^{m'} \notag \\
&   \le \tilde{g}(\ell+1)^{-\alpha}  \sum_{m'=0}^\infty  (m'+1)^2 e^{-m'} 
\le 6 \tilde{g}(\ell+1)^{-\alpha},
\end{align}
where we use $|t|\le 1/(2e\tilde{g})$ and $\sum_{m'=0}^\infty  (m'+1)^2 e^{-m'} = 5.41562\cdots < 6$. 
Because of $\ell=\diam(\Lambda_w)$, we obtain the inequality~\eqref{supp_first_step_inequality}.

\subsubsection{Proof of Lemma~\ref{supp_lem:2point_include_sum}}

In order to characterize the subset-subset connections, we first introduce a set $\mathcal{G}_m$ of graph structures (Fig.~\ref{supp_connection_graph}).
Each of the graph $G=(V,E) \in \mathcal{G}_m$ ($|V|=m+1$) is constructed recursively as follow: 
the first vertex has a node with the vertex $0$.  
The second vertex has a node with vertex $0$ or $1$. 
By repeating this procedure,  $j$th vertex has a node with vertex in $\{0,1,2,\ldots, j-1\}$.
In this construction, we define $E$ as the node set as $\{(1,\ed_1), (2,\ed_2) , (3,\ed_3), \cdots (m,\ed_m)\}$ ($\ed_j \in \{0,1,2,\ldots, j-1\}$, $\ed_j < j$), where $\ed_1=0$.  
In Fig.~\ref{supp_connection_graph} (a), we show one example of $m=6$. In Fig.~\ref{supp_connection_graph} (b), we show all the patterns of the graph in $\mathcal{G}_3$.
We notice that the number of graph in $\mathcal{G}_m$ is equal to $m!$, namely $|\mathcal{G}_m|=m!$.

For a fixed graph $G\in \mathcal{G}_m$, we define the set $\Omega_G$ of $w$ as follows:
\begin{align}
\Omega_G := \{w \in \Omega_m | Z_1 \cap Z_0 \neq \emptyset, Z_2 \cap Z_{\ed_2} \neq \emptyset ,\ldots  Z_m \cap Z_{\ed_m} \neq \emptyset \} . 
\end{align}
We note that in the above restrictions, a subset $Z_j$ must connect to the subset $Z_{\ed_j}$ ($Z_j\cap Z_{\ed_j}\neq \emptyset$) which is connected to $Z_j$ on the graph $G$, but it does not necessarily mean $Z_j \cap Z_{j'} = \emptyset$ for $j' \neq \ed_j$. 
Then, all the strings $w\in \Omega^\ast_m$ satisfying the condition~\eqref{supp_condition_subset_Z_j} can be (over)counted by considering $w\in \Omega_G$ for all $G\in \mathcal{G}_m$, namely
\begin{align}
\Omega^\ast_m = \bigcup_{G\in\mathcal{G}_m} \Omega_G .
\end{align}
We thus reduce
\begin{align}
\label{supp_graph_summation_decomp}
\sum_{\substack{ w\in \Omega^\ast_{m} \\  \overline{\Lambda}_{w} \ni \{i,i'\}}} |J_{w}|    
\le \sum_{G\in \mathcal{G}_{m}}\sum_{\substack{ w\in \Omega_{G} \\  \overline{\Lambda}_{w} \ni \{i,i'\} }} |J_{w}| .
\end{align}
In the above summation, the same string $w$ may be counted in $\Omega_{G_1}$ and $\Omega_{G_2}$ for different $G_1$ and $G_2$.

Let us consider a fixed $G$.
Then, in every string $w \in \Omega_G$ satisfying $\Lambda_{w} \ni \{i,i'\}$, there exist two subsets $Z_s$ and $Z_{s'}$ ($s,s'=0,1,2,\ldots,m$) such that $Z_s \ni i$ and $Z_{s'}\ni i'$.
We need to consider all the combinations of $\{s,s'\}$ (i.e., $\multiset{m+1}{2}$ patterns in total), but we here estimate the contribution from one of them. 
For a fixed $\{s,s'\}$, we label $Z_s$ and $Z_{s'}$ as $Z_{s_1}$ and $Z_{s_l}$, where $(l-1)$ is the path length to connect $Z_s$ and $Z_s'$ on the graph $G$ (see Fig.~\ref{supp_fig_w'_sum}). 
We then decompose $w$ into $w_1$ and $w_2$, where $w_1=((Z_{s_1},q_{s_1}),(Z_{s_2},q_{s_2}), \ldots, (Z_{s_l},q_{s_l}))$ and $w_2=w\setminus w_1$.
In the graph $G$, we have $\ed_{s_2}=s_1$, $\ed_{s_3}=s_2$, $\ldots$ ,$\ed_{s_{l-1}}=s_l$ and $Z_{s_1} \ni i$, $Z_{s_l} \ni i'$. 

\begin{figure}[tt]
\centering
\includegraphics[clip, scale=0.7]{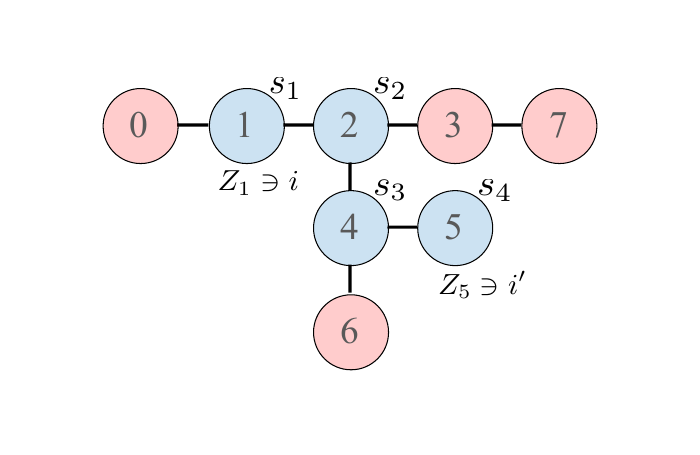}
\caption{Decomposition of $w\in \Omega_{G}$ into two pieces $w_1$ and $w_2$. 
In the above picture, we consider the case that $Z_1 \ni i$ and $Z_5\ni i'$. 
The string $w_1$ consists of the elements which connect from $Z_1$ to $Z_5$ [i.e., $w_1=((Z_1,q_1),(Z_2,q_2),(Z_4,q_4),(Z_5,q_5))$], and the other elements are included in $w_2$ [i.e., $w_2=((Z_0,q_0),(Z_3,q_3),(Z_6,q_6),(Z_7,q_7))$].
}
\label{supp_fig_w'_sum}
\end{figure}

We then estimate the summation with respect to $w\in \Omega_{G}$ for a fixed $G$ such that $Z_{s_1} \ni i$, $Z_{s_l} \ni i'$:
\begin{align}
\label{supp_summation_w'_for_fixed_G'_i_i'}
\sum_{\substack{ w\in \Omega_{G} \\  Z_{s_1} \ni i, Z_{s_l} \ni i'}} |J_{w}|  .
\end{align}
We first take the summation with respect to $w_2$.  
The contribution from the $w_2$ comes from a summation like 
\begin{align}
\sum_{\substack{(Z_j,q_j): Z_j \cap Z_{\ed_j}\neq 0}} |J_{Z_j,q_j} |  .
\end{align}
In order to estimate the upper bound, we use the following general upper bound which is derived from the inequality~\eqref{supp_g_extensive}: 
\begin{align}
\label{supp_eq:extensive_estimation}
\sum_{\substack{(Z_j,q_j): Z_j \cap Z_{\ed_j}\neq 0}} |J_{Z_j,q_j} |  \le \sum_{i\in Z_{\ed_j}} \sum_{\substack{(Z_j,q_j): Z_j \ni i}} |J_{Z_j,q_j} |  
\le g |Z_{\ed_j}| \le gk.
\end{align}
By using the above inequality, the summation~\eqref{supp_summation_w'_for_fixed_G'_i_i'} reduces to  
  \begin{align}
  \label{supp_ineq:the summation for w'_1 is}
&\sum_{\substack{ w\in \Omega_{G} \\  Z_{s_1} \ni i, Z_{s_l} \ni i'}} |J_{w}|  \notag \\
&\le (gk)^{|w_2|} \sum_{\substack{(Z_{s_1},q_{s_1})\\ Z_{s_1} \ni i}} |J_{Z_{s_1},q_{s_1}} |   
\sum_{\substack{ (Z_{s_2},q_{s_2}) \\ Z_{s_2} \cap Z_{s_1}\neq \emptyset}}   |J_{Z_{s_2},q_{s_2}} | \cdots
\sum_{\substack{ (Z_{s_{l-1}},q_{s_{l-1}}) \\ Z_{s_{l-1}} \cap Z_{s_{l-2}}\neq \emptyset}}   |J_{Z_{s_{l-1}},q_{s_{l-1}}} |
\sum_{\substack{ (Z_{s_l},q_{s_l}) \\ Z_{s_l} \cap Z_{s_{l-1}}\neq \emptyset , Z_{s_l} \ni i'}}   |J_{Z_{s_l},q_{s_l}} |  .
\end{align}
Remember that $((Z_{s_j},q_{s_j}))_{j=1}^l$ now corresponds to $w_1$. 

In order to estimate the summation with respect to $w_1$, we use a more refined analysis.
First, for the summation of $Z_{s_1}$ and $Z_{s_2}$, we use the summation reduction as 
 \begin{align}
\sum_{\substack{(Z_{s_1},q_{s_1})\\ Z_{s_1} \ni i }} |J_{Z_{s_1},q_{s_1}} |   
\sum_{\substack{ (Z_{s_2},q_{s_2}) \\ Z_{s_2} \cap Z_{s_1}\neq \emptyset}}   |J_{Z_{s_2},q_{s_2}} |  
\le \sum_{i_1\in \Lambda}\sum_{\substack{Z_{s_1},q_{s_1}\\ Z_{s_1} \ni \{i,i_1\}}}  |J_{Z_{s_1},q_{s_1}} |   
\sum_{\substack{Z_{s_2},q_{s_2} \\ Z_{s_2} \ni i_1}}   |J_{Z_{s_2},q_{s_2}} |  .
\end{align}
By iteratively using the above reductions, the summation in~\eqref{supp_ineq:the summation for w'_1 is} is bounded from above by
 \begin{align}
&\sum_{i_1,i_2,\ldots, i_{l-1}\in \Lambda}
\sum_{\substack{Z_{s_1},q_{s_1}\\ Z_{s_1} \ni \{i,i_1\}}}  |J_{Z_{s_1},q_{s_1}} |   
\sum_{\substack{Z_{s_2},q_{s_2} \\ Z_{s_2} \ni \{i_1,i_2\}}}    |J_{Z_{s_2},q_{s_2}} |  
\cdots 
\sum_{\substack{Z_{s_{l-1}},q_{s_{l-1}}\\ Z_{s_{l-1}} \ni \{i_{l-2},i_{l-1}\}}}  |J_{Z_{s_{l-1}},q_{s_{l-1}}} |
\sum_{\substack{Z_{s_{l}},q_{s_{l}}\\ Z_{s_{l}} \ni \{i_{l-1},i'\}}}  |J_{Z_{s_l},q_{s_l}} |   \notag \\
\le & J^l \sum_{i_1,i_2,\ldots, i_{l-1}\in \Lambda} (\dist_{i,i_1}+1)^{-\alpha} (\dist_{i_1,i_2}+1)^{-\alpha} \cdots (\dist_{i_{l-2},i_{l-1}}+1)^{-\alpha}( \dist_{i_{l-1},i'}+1)^{-\alpha},
\end{align}
where we use the power-law decay of the interactions for each of $\{ |J_{Z_{s_{j}},q_{s_{j}}} | \}_{j=1}^l$
Following Ref.~\cite[Inequality~(2.5)]{ref:Hastings2006-ExpDec}, we here utilize the upper bound of 
\begin{align}
\sum_{i_0 \in \Lambda}  (\dist_{i,i_0}+1)^{-\alpha}  (\dist_{i_0,i'}+1)^{-\alpha}  \le \lambda (\dist_{i,i'}+1)^{-\alpha}   ,
\end{align}
where $\lambda>1$ is a constant of $\orderof{1}$ as long as $\alpha>D$.
We thus obtain 
 \begin{align}
&J^l \sum_{i_1,i_2,\ldots, i_{l-1}\in \Lambda}(\dist_{i,i_1}+1)^{-\alpha} (\dist_{i_1,i_2}+1)^{-\alpha}  \cdots (\dist_{i_{l-2},i_{l-1}}+1)^{-\alpha}( \dist_{i_{l-1},i'}+1)^{-\alpha}   \notag \\
&\le J^l \lambda^{l-1} (\dist_{i,i'}+1)^{-\alpha}\le (\lambda J)^l (\dist_{i,i'}+1)^{-\alpha}.
\end{align}
By combining the above inequalities together,  the summation~\eqref{supp_summation_w'_for_fixed_G'_i_i'} is upper-bounded as 
  \begin{align}
\sum_{\substack{ w\in \Omega_{G} \\  Z_{s_1} \ni i, Z_{s_l} \ni i'}} |J_{w}|  \le [\max(gk, \lambda J)]^{m+1} (\dist_{i,i'}+1)^{-\alpha}.
\end{align}
The number of combinations of $Z_s$ and $Z_{s'}$ such that $Z_s \ni i$ and $Z_{s'}\ni i'$ is given by
  \begin{align}
\multiset{m+1}{2} = \binom{m+2}{2}  \le (m+1)^2.
\end{align}
Therefore, we finally obtain 
  \begin{align}
\sum_{\substack{ w\in \Omega_{G} \\  \overline{\Lambda}_{w} \ni \{i,i'\}}} |J_{w}|  \le (m+1)^2\tilde{g}^{m+1}(\dist_{i,i'}+1)^{-\alpha} ,
\end{align}
where we use the definition $\tilde{g}:=\max(gk, \lambda J)] $. 
Because the number of graphs such that $G\in \mathcal{G}_m$ is equal to $m!$, we obtain the inequality~\eqref{supp_2point_include_sum_upp}.  
This completes the proof. $\square$


 {~}

\hrulefill{\bf [ End of Proof of Lemma~\ref{supp_lem:2point_include_sum}] }

{~}


\subsection{Estimation of the summation with respect to $w$: Proof of \eqref{supp_aim_eq:estimate the summation with respect to w}. }

We here consider the summation with respect to $w$ as 
 \begin{align}
 \label{supp_summation_w_diam_taking_into_account}
\sum_{\substack{ w\in \Omega^\ast_m :Z_0\in \Set_{\le r/2}\\  \Lambda_w\ni i}} \frac{| J_w |}{[\diam(\Lambda_w)+1]^\alpha} .
 \end{align}
Now, the difficulty lies in the fact that we need to take $\diam(\Lambda_w)$ into account. 
Intuitively, in the above summation, we may have $\diam(\Lambda_w) =\orderof{r}$ from $\diam(\overline{\Lambda}_w) > r/2$, but it is not always true since the form of $\Lambda_w$ strongly depends on properties of the string $w$. 


In order to derive the upper bound for \eqref{supp_summation_w_diam_taking_into_account}, we prove the following Lemma (see Sec.~\ref{supp_sec:Proof of Lemma_lemma_decomp_path} for the proof):
\begin{lemma} \label{supp_lemma_decomp_path}
Let $w \in \Omega^\ast_m$ be an arbitrary string such that $q_w \neq 0$. 
Then, there exists a decomposition of $w$ to $w_1$ and $w_2$ each of which has the following properties:
\begin{enumerate} \label{supp_efefjei}
\item{} For an arbitrary element in $w_1$ ($w_2$), there exists a path which connect an arbitrary element to $Z_0$ via the subsets in $w_1$ ($w_2$).
\item{} The subset $\Lambda_w$ satisfies $\overline{\Lambda}_{w_1} \cap \Lambda_w \neq \emptyset$ and  $\overline{\Lambda}_{w_2} \cap \Lambda_w \neq \emptyset$, 
where  $\overline{\Lambda}_w$ has been defined in Eq.~\eqref{supp_def:tilde_V___w}. 
\end{enumerate}
\end{lemma}
\noindent 
This lemma implies in order to make $\diam(\Lambda_w)\le \ell$, there should exist the following two paths in $w$: $Z_0 \to i$ ($i \in X[r]^\co$)  and $Z_0\to i' $ with $\dist_{i,i'} \le \ell$. 
For $\Lambda_w\ni i$ ($\dist_{i,X}\ge r$), without loss of generality, we choose the string $w_1$ and $w_2$ such that $w_1$ includes the element $(Z_0,q_0)$ and 
$\overline{\Lambda}_{w_2}$ includes the site $i$ (i.e., $\overline{\Lambda}_{w_2}\ni i$). 

\begin{figure}[tt]
\centering
\includegraphics[clip, scale=0.5]{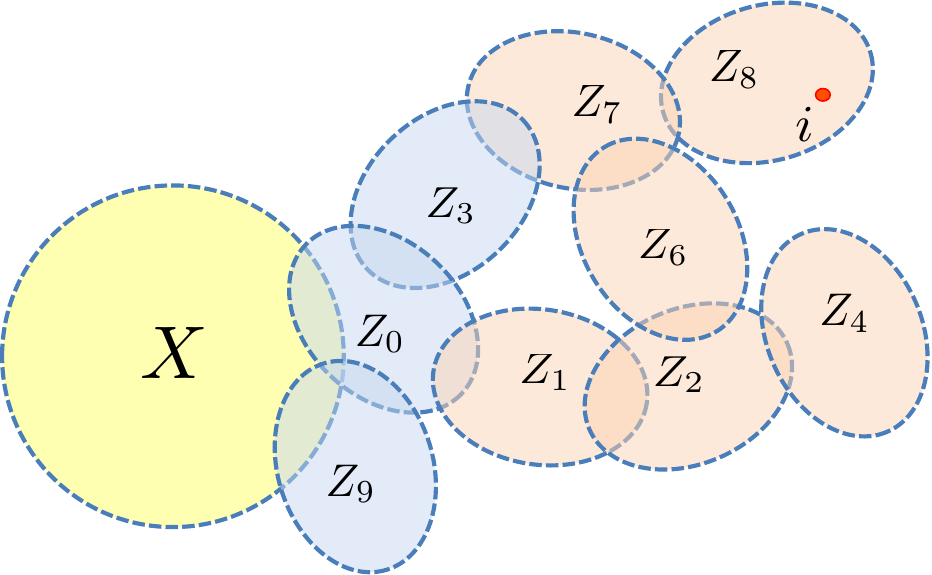}
\caption{The decomposition of string $w$ to $w_1$ and $w_2$ following Lemma~\ref{supp_lemma_decomp_path}.
Here, each of $w_1$ and $w_2$ satisfies the properties 1 and 2 in Lemma~\ref{supp_lemma_decomp_path}. 
In our choice, the string $w_1=((Z_0,q_0),(Z_3,q_3),(Z_9,q_9))$ includes the element $(Z_0,q_0)$, while $w_2=((Z_1,q_1),(Z_2,q_2),(Z_4,q_4),(Z_6,q_6),(Z_7,q_7),(Z_8,q_8))$ necessarily includes the site $i$.
}
\label{supp_fig_w1_w2_decomp}
\end{figure}

By using Lemma~\ref{supp_lemma_decomp_path}, we decompose the summation with respect to $w\in \Omega^\ast_m$ such that $\Lambda_w\ni i$ as follows:
\begin{align}
 \label{supp_eq:estimate the summation with respect to w_1_w_2_00}
\sum_{\substack{ w\in \Omega^\ast_m :Z_0\in \Set_{\le r/2}\\  \Lambda_w\ni i}} \frac{| J_w |}{[\diam(\Lambda_w)+1]^\alpha} 
&\le \sum_{m_1=0}^{m-1} \binom{m}{m_1}  
 \sum_{\substack{ w_1\in \Omega^\ast_{m_1} \\  Z_0\in \Set_{\le r/2}}}
 \sum_{  \substack{w_2\in \Omega^\ast_{m_2-1} \\   \overline{\Lambda}_{w_2} \cap Z_0\neq \emptyset , \overline{\Lambda}_{w_2}\ni i  } }  \frac{| J_{w_1} | \cdot | J_{w_2} | }{[\min_{i'\in \overline{\Lambda}_{w_1}}(\dist_{i,i'})+1]^\alpha}  ,
 \end{align}
where we set $|w_1|=m_1+1$ and $|w_2|=m_2+1$ ($m_1+m_2=m$), and for each of $w$,  we use the inequality of
  \begin{align}
\diam(\Lambda_w) \ge \max_{i_1, i_2\in \Lambda_w} (\dist_{i_1,i_2}) \ge \min_{i'\in \overline{\Lambda}_{w_1}}  (\dist_{i,i'}).
\end{align}
We reduce the inequality~\eqref{supp_eq:estimate the summation with respect to w_1_w_2_00} to 
\begin{align}
 \label{supp_eq:estimate the summation with respect to w_1_w_2}
\sum_{\substack{ w\in \Omega^\ast_m :Z_0\in \Set_{\le r/2}\\  \Lambda_w\ni i}} \frac{| J_w |}{[\diam(\Lambda_w)+1]^\alpha} 
&\le \sum_{l=-\infty}^{r/2} \sum_{i_1 \in (\partial X)_l} \sum_{m_1=0}^{m-1} \binom{m}{m_1}  
  \sum_{  \substack{w_2\in \Omega^\ast_{m_2-1} \\    \overline{\Lambda}_{w_2}\ni \{i_1, i\}  } } 
  \sum_{\substack{ w_1\in \Omega^\ast_{m_1} \\  Z_0\ni i_1}}
 \frac{| J_{w_1} | \cdot | J_{w_2} | }{[\min_{i'\in \overline{\Lambda}_{w_1}}(\dist_{i,i'})+1]^\alpha} .
 \end{align}


We first estimate the summation with respect to $w_1$. We separate the summation to the cases of $\min_{i'\in \overline{\Lambda}_{w_1}}(\dist_{i,i'}) \le (\dist_{i,i_1}/2)$ and $\min_{i'\in \overline{\Lambda}_{w_1}}(\dist_{i,i'}) > (\dist_{i,i_1}/2)$:
 \begin{align}
 \label{supp_eq:estimate the summation with respect to w_2}
\sum_{\substack{ w_1\in \Omega^\ast_{m_1} \\  Z_0\ni i_1}} \frac{| J_{w_1} |}{[\min_{i'\in \overline{\Lambda}_{w_1}}(\dist_{i,i'}+1)]^\alpha} 
\le  \sum_{x=0}^{\dist_{i,i_1}/2}\sum_{i'\in \Lambda: \dist_{i,i'} = x}\sum_{\substack{ w_1\in \Omega^\ast_{m_1} \\  Z_0\ni i_1, \overline{\Lambda}_{w_1}\ni i'}}
 \frac{| J_{w_1} |}{(x+1)^\alpha} + \sum_{\substack{ w_1\in \Omega^\ast_{m_1} \\  Z_0\ni i_1}}  \frac{| J_{w_1} |}{(\dist_{i,i_1}/2+1)^\alpha} ,
 \end{align}
where we use the inequality of
 \begin{align}
\sum_{\substack{ w_1\in \Omega^\ast_{m_1} \\  Z_0\ni i_1, \min_{i'\in \overline{\Lambda}_{w_1}}(\dist_{i,i'})=x}} \frac{| J_{w_1} |}{[\min_{i'\in \overline{\Lambda}_{w_1}}(\dist_{i,i'}+1)]^\alpha}  \le \sum_{i'\in \Lambda: \dist_{i,i'} = x}\sum_{\substack{ w_1\in \Omega^\ast_{m_1} \\  Z_0\ni i_1, \overline{\Lambda}_{w_1}\ni i'}} \frac{| J_{w_1} |}{(x+1)^\alpha} .
 \end{align}
We first consider the second term. 
By using the decomposition of \eqref{supp_graph_summation_decomp} and the inequality~\eqref{supp_eq:extensive_estimation}, we have 
  \begin{align}
   \label{supp_summation_w_2_second_term}
\sum_{\substack{ w_1\in \Omega^\ast_{m_1} \\  Z_0\ni i_1}}  \frac{| J_{w_1} |}{(\dist_{i,i_1}/2+1)^\alpha}
\le \frac{m_1! g(gk)^{m_1}}{(\dist_{i,i_1}/2+1)^{\alpha}} \le \frac{m_1! (gk)^{m_1+1}}{(\dist_{i,i_1}/2+1)^{\alpha}}.
 \end{align} 
 For the first term in \eqref{supp_eq:estimate the summation with respect to w_2}, we utilize Lemma~\ref{supp_lem:2point_include_sum} as follows:
  \begin{align}
\sum_{\substack{ w_1\in \Omega^\ast_{m_1} \\  Z_0\ni i_1, \overline{\Lambda}_{w_1}\ni i'}}
 \frac{| J_{w_1} |}{(x+1)^\alpha}
\le  \frac{1}{(x+1)^\alpha} \sum_{ \substack{ w_1\in \Omega^\ast_{m_1} \\ \overline{\Lambda}_{w_1}\ni \{i_1,i'\}} } | J_{w_1} | 
\le \frac{m_1! (m_1+1)^2\tilde{g}^{m_1+1}}{(x+1)^{\alpha} (\dist_{i_1,i'} +1)^{\alpha}}  ,
 \end{align}
which yields
 \begin{align}
 \label{supp_summation_w_2_simplify}
\sum_{x=0}^{\dist_{i,i_1}/2}\sum_{i'\in \Lambda: \dist_{i,i'} = x}\sum_{\substack{ w_1\in \Omega^\ast_{m_1} \\  Z_0\ni i_1, \overline{\Lambda}_{w_1}\ni i'}}
 \frac{| J_{w_1} |}{(x+1)^\alpha} 
 \le m_1! (m_1+1)^2\tilde{g}^{m_1+1}  \sum_{x=0}^{\dist_{i,i_1}/2}\sum_{i'\in \Lambda: \dist_{i,i'} = x} 
(x+1)^{-\alpha} (\dist_{i_1,i'} +1)^{-\alpha}      .
 \end{align}

 \begin{figure}[tt]
\centering
\includegraphics[clip, scale=0.5]{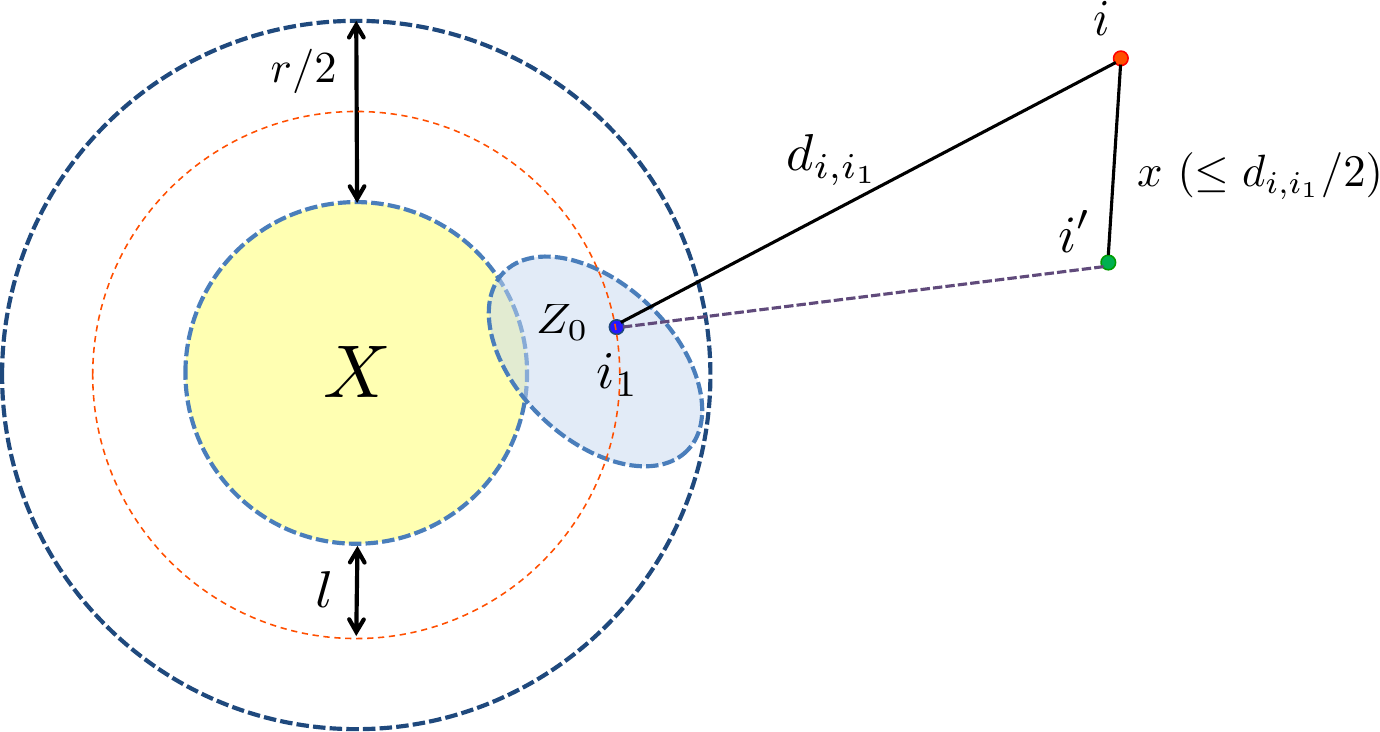}
\caption{Schematic picture of the positions of $Z_0$, $i_1$, $i$ and $i'$. Here, $Z_0$ includes a site $i_1$ on $(\partial X)_l$ (red dot region). 
The distance between the site $i$ and $i'$ is defined as $x$ and is smaller than or equal to $\dist_{i,i_1}/2$.
}
\label{supp_fig_triangle_pos}
\end{figure}
 
Here, the lengths $\dist_{i,i'}$ ($=x$) satisfies $\dist_{i,i'} \le \dist_{i,i_1}/2$ (see Fig.~\ref{supp_fig_triangle_pos}), and hence the triangle inequality gives 
 \begin{align}
\dist_{i_1,i'} \ge \dist_{i_1,i} - \dist_{i,i'}   \ge \frac{\dist_{i,i_1}}{2}.
 \end{align}
We thus reduce the inequality~\eqref{supp_summation_w_2_simplify} to
 \begin{align}
 \label{supp_summation_w_2_simplify_final}
\sum_{x=0}^{\dist_{i,i_1}/2}\sum_{i'\in \Lambda: \dist_{i,i'} = x}\sum_{\substack{ w_1\in \Omega^\ast_{m_1} \\  Z_0\ni i_1, \overline{\Lambda}_{w_1}\ni i'}}
 \frac{| J_{w_1} |}{(x+1)^\alpha} 
& \le  \frac{m_1! (m_1+1)^2\tilde{g}^{m_1+1} }{(\dist_{i,i_1}/2+1)^{\alpha}} \sum_{x=0}^{\dist_{i,i_1}/2}\sum_{i'\in \Lambda: \dist_{i,i'} = x} (x+1)^{-\alpha}  \notag \\
 &\le \frac{m_1! (m_1+1)^2\tilde{g}^{m_1+1}}{(\dist_{i,i_1}/2+1)^{\alpha}} \br{ 1 +\frac{\gamma}{\alpha-D}},
 \end{align}
where we use a similar inequality to \eqref{supp_ineq_sum_decay_func}. 
By applying the inequalities~\eqref{supp_summation_w_2_second_term} and \eqref{supp_summation_w_2_simplify_final} to \eqref{supp_eq:estimate the summation with respect to w_2}, we obtain 
 \begin{align}
 \label{supp_eq:estimate the summation with respect to w_2_final_form}
\sum_{\substack{ w_1\in \Omega^\ast_{m_1} \\  Z_0\ni i_1}} \frac{| J_{w_1} |}{[\min_{i'\in \overline{\Lambda}_{w_1}}(\dist_{i,i'}+1)]^\alpha} 
&\le \frac{m_1! (gk)^{m_1+1}}{(\dist_{i,i_1}/2+1)^{\alpha}} +  \frac{m_1! (m_1+1)^2\tilde{g}^{m_1+1}}{(\dist_{i,i_1}/2+1)^{\alpha}} \br{ 1 +\frac{\gamma}{\alpha-D}} \notag \\
&\le \frac{ c_2m_1! (m_1+1)^2\tilde{g}^{m_1+1}}{(\dist_{i,i_1}+1)^{\alpha}}  ,
 \end{align}
where we use  $\tilde{g}\ge gk$, and defined $c_2$ as in Eq.~\eqref{supp_definition_of_c_2}.

The remaining task is to take the summation with respect to $w_2$ in the inequality \eqref{supp_eq:estimate the summation with respect to w_1_w_2}.
By using Lemma~\ref{supp_lem:2point_include_sum}, we immediately obtain 
\begin{align}
 \label{supp_eq:estimate the summation with respect to w_1_final_form}
 \sum_{  \substack{w_2\in \Omega^\ast_{m_2-1} \\    \overline{\Lambda}_{w_2}\ni \{i_1, i\}  } }   | J_{w_2} | 
&\le \frac{ (m_2-1)! m_2^2\tilde{g}^{m_2}}{(\dist_{i,i_1}+1)^{\alpha}} =\frac{ m_2! m_2 \tilde{g}^{m_2}}{(\dist_{i,i_1}+1)^{\alpha}}  .
 \end{align}
 By combining the inequalities~\eqref{supp_eq:estimate the summation with respect to w_2_final_form} and~\eqref{supp_eq:estimate the summation with respect to w_1_final_form},
 the inequality~\eqref{supp_eq:estimate the summation with respect to w_1_w_2} reduces to
 \begin{align}
\sum_{\substack{ w\in \Omega^\ast_m :Z_0\in \Set_{\le r/2}\\  \Lambda_w\ni i}} \frac{| J_w |}{[\diam(\Lambda_w)+1]^\alpha} 
&\le \sum_{l=-\infty}^{r/2} \sum_{i_1 \in (\partial X)_l}  \sum_{m_1=0}^{m-1} \binom{m}{m_1}  \frac{c_2  m_1! (m_1+1)^2\tilde{g}^{m_1+1} \cdot  m_2! m_2 \tilde{g}^{m_2}}{(\dist_{i,i_1}+1)^{2\alpha}} 
\notag\\
&=c_2 m! \tilde{g}^{m+1} \sum_{l=-\infty}^{r/2} \sum_{i_1 \in (\partial X)_l}  \frac{1}{(\dist_{i,i_1}+1)^{2\alpha}}  \sum_{m_1=1}^m(m_1+1)^2 m_2  \notag \\
&\le c_2 m!  (m+1)^4\tilde{g}^{m+1} \sum_{l=-\infty}^{r/2} \sum_{i_1 \in (\partial X)_l}  \frac{1}{(\dist_{i,i_1}+1)^{2\alpha}}.
 \end{align}
 This completes the proof of the inequality~\eqref{supp_aim_eq:estimate the summation with respect to w}.

\subsubsection{Proof of Lemma~\ref{supp_lemma_decomp_path}} \label{supp_sec:Proof of Lemma_lemma_decomp_path}

We first focus on the following fact.
For arbitrary $X$ and $Y$ ($X\cap Y \neq \emptyset$), let us consider the commutator between the operators $\Pau_{X,q}$ and $\Pau_{Y,q'}$ as
\begin{align}
[\Pau_{X,q}, \Pau_{Y,q'} ]  \propto \Pau_{Z,q''}. 
\end{align}
Then, if $q''\neq 0$, we  obtain 
\begin{align}
X \cap Z \neq \emptyset, \quad Y \cap Z \neq \emptyset  .\label{supp_First_step_cond}
\end{align}

 
We prove the statement in Lemma~\ref{supp_lemma_decomp_path} by induction method. 
For this purpose, we define $w^{(p)}$ ($p\le m$) as $w^{(p)}:=((Z_j,q_j))_{j=0}^p$. 
First, for $p=1$, we choose $w_1^{(1)}=((Z_0,q_0))$ and $w_2^{(1)}=((Z_1,q_1))$, and then the property~1 in Lemma~\ref{supp_lemma_decomp_path} is trivially satisfied.
Also, from Eq.~\eqref{supp_First_step_cond}, we have $\Lambda_{w^{(1)}} \cap Z_0 \neq \emptyset$ and $\Lambda_{w^{(1)}} \cap Z_1 \neq \emptyset$, which gives the property~2.

In general $p$, we assume the decomposition of $w^{(p)}=w_1^{(p)}\oplus w_2^{(p)}$ with the desired properties and consider the case of $p+1$. 
We here consider the cases of 
\begin{align}
\label{supp_case_1_general_p_proof}
\overline{\Lambda}_{w_1} \cap Z_{p+1} = \emptyset,\quad {\rm or} \quad \overline{\Lambda}_{w_2} \cap Z_{p+1} = \emptyset
\end{align}
and 
\begin{align}
\label{supp_case_2_general_p_proof}
\overline{\Lambda}_{w_1} \cap Z_{p+1} \neq \emptyset,\quad {\rm and} \quad \overline{\Lambda}_{w_2} \cap Z_{p+1} \neq \emptyset
\end{align}
separately. 

{~}\\

{\bf [Case of~\eqref{supp_case_1_general_p_proof}]} \\
Let us consider the case of $\overline{\Lambda}_{w_1} \cap Z_{p+1} = \emptyset$ (i.e., $\overline{\Lambda}_{w_2} \cap Z_{p+1} \neq \emptyset$). 
We here choose $w_1^{(p+1)} = w_1^{(p)}$ and $w_2^{(p+1)}=w_2^{(p)} \oplus (Z_{p+1},q_{p+1})$.
This choice trivially satisfies the property~1 under the assumption that the decomposition $w^{(p)}=w_1^{(p)}\oplus w_2^{(p)}$ satisfies the property~1.
On the second property, we first set $\Lambda_1=\overline{\Lambda}_{w_1^{(p)}} \cap \Lambda_{w^{(p)}}$ and $\Lambda_2=\Lambda_{w^{(p)}} \setminus \Lambda_1$. 
Note that $\Lambda_2 \in \overline{\Lambda}_{w_2^{(p)}}$ because of $\Lambda_{w^{(p)}}  \subseteq \overline{\Lambda}_{w_1^{(p)}} \cup \overline{\Lambda}_{w_2^{(p)}}$.

We then decompose
\begin{align}
\label{supp_decomp_p_w}
\Pau_{\Lambda_{w^{(p)}},q_{w^{(p)}}}  = \Pau_{\Lambda_1,q_1} \otimes \Pau_{\Lambda_2,q_2}.
\end{align}
Because $\Lambda_1\cap Z_{p+1}=\emptyset$ from $\overline{\Lambda}_{w_1} \cap Z_{p+1} = \emptyset$, we have 
\begin{align}
[ \Pau_{\Lambda_{w^{(p)}},q_{w^{(p)}}}, \Pau_{Z_{p+1},q_{p+1}}]  
&= \Pau_{\Lambda_1,q_1} \otimes \left[\Pau_{\Lambda_2,q_2} , \Pau_{Z_{p+1},q_{p+1}}\right] \propto  \Pau_{\Lambda_1,q_1} \otimes \Pau_{\Lambda'_2,q_2}   ,
\end{align}
where $\Lambda'_2 \subseteq \overline{\Lambda}_{w_2^{(p+1)}}$ and $\Lambda'_2\neq \emptyset$ from Eq.~\eqref{supp_First_step_cond}\footnote{
We note that $\Lambda_2 \neq \emptyset$ is ensured because of $q_w \neq 0$. If $\Lambda_2=\emptyset$, we have 
$\Pau_{\Lambda_1,q_1} \otimes\left[\Pau_{\Lambda_2,q_2} , \Pau_{Z_{p+1},q_{p+1}}\right] = \Pau_{\Lambda_1,q_1} \otimes\left[1 , \Pau_{Z_{p+1},q_{p+1}}\right]  =0$, which yields $q_{w^{(p+1)}}=0$. We notice that  $q_{w^{(p)}}\neq 0$ is ensured for arbitrary $p=1,2, \ldots ,m$ because of the condition $q_w=q_{w^{(m)}} \neq 0$.}. 
Therefore, we obtain $\Lambda_1\subseteq \overline{\Lambda}_{w_1^{(p+1)}}$ and $\Lambda'_2\subseteq \Lambda_{w_2^{(p+1)}}$, which yields the property~2.  
 
{~}\\

{\bf [Case of~\eqref{supp_case_2_general_p_proof}]} \\
We first consider the case 
\begin{align}
\overline{\Lambda}_{w_1^{(p)}} \cap \Lambda_{w^{(p+1)}} \neq \emptyset  \quad {\rm and} \quad \overline{\Lambda}_{w_2^{(p)}} \cap \Lambda_{w^{(p+1)}} \neq \emptyset .
\end{align}
In this case, by choosing $w_1^{(p+1)} = w_1^{(p)}$ and $w_2^{(p+1)}=w_2^{(p)} \oplus (Z_{p+1},q_{p+1})$, we can trivially obtain the properties~2.
Also, the property~1 is ensured by the condition~\eqref{supp_case_2_general_p_proof}.
Hence, we need to prove the case where 
\begin{align}
\overline{\Lambda}_{w_1^{(p)}} \cap \Lambda_{w^{(p+1)}} = \emptyset  \quad {\rm or} \quad \overline{\Lambda}_{w_2^{(p)}} \cap \Lambda_{w^{(p+1)}} = \emptyset .
\end{align}

In the following, let us consider the former case (the latter case can be treated in the same way).
We then choose $w_1^{(p+1)} = w_1^{(p)} \oplus (Z_{p+1},q_{p+1})$ and $w_2^{(p+1)}=w_2^{(p)}$.
Then, the property~1 is obtained by the condition~\eqref{supp_case_2_general_p_proof}.
In order to prove the property~2, we first adopt the same notation as Eq.~\eqref{supp_decomp_p_w}, and consider 
\begin{align}
[ \Pau_{\Lambda_1,q_1} \otimes \Pau_{\Lambda_2,q_2}, \Pau_{Z_{p+1},q_{p+1}}]  \propto \Pau_{\Lambda_{w^{(p+1)}},q_{w^{(p+1)}}}
\end{align}
with $q_{w^{(p+1)}}\neq 0$.
Then, the condition $\overline{\Lambda}_{w_1^{(p)}} \cap \Lambda_{w^{(p+1)}} = \emptyset $ implies 
\begin{align}
\Lambda_1 \cap \Lambda_{w^{(p+1)}} \subseteq  \overline{\Lambda}_{w_1^{(p)}} \cap \Lambda_{w^{(p+1)}}= \emptyset ,
\end{align}
where the first relation comes from $\Lambda_1 \subseteq \overline{\Lambda}_{w_1^{(p)}}$. 
On the other hand, from Eq.~\eqref{supp_First_step_cond}, we have $(\Lambda_1\cup \Lambda_2) \cap \Lambda_{w^{(p+1)}} \neq \emptyset$, and hence 
\begin{align}
\Lambda_2 \cap \Lambda_{w^{(p+1)}} \neq \emptyset, 
\end{align}
which implies $\overline{\Lambda}_{w_2^{(p+1)}} \cap \Lambda_{w^{(p+1)}} \neq \emptyset$.
Also, from Eq.~\eqref{supp_First_step_cond}, we have $Z_{p+1} \cap \Lambda_{w^{(p+1)}} \neq \emptyset$, and hence we have 
$\overline{\Lambda}_{w_1^{(p+1)}} \cap \Lambda_{w^{(p+1)}} \neq \emptyset$ because of $Z_{p+1}\subseteq \overline{\Lambda}_{w_1^{(p+1)}}=\overline{\Lambda}_{w_1^{(p)}} \cup Z_{p+1}$.
We thus prove the property~2.

We thus prove the properties of 1 and 2 for both of the cases~\eqref{supp_case_1_general_p_proof} and \eqref{supp_case_2_general_p_proof} for $p+1$. 
This completes the proof of Lemma~\ref{supp_lemma_decomp_path}. $\square$


 {~}

\hrulefill{\bf [ End of Proof of Lemma~\ref{supp_lemma_decomp_path}] }

{~}




\end{widetext}

\end{document}